\documentstyle[11pt]{article}
\begin{document}
\title{{\bf Supervisory Control of Fuzzy Discrete Event Systems: A Formal Approach}\thanks{This work was supported in part by the National
Natural Science Foundation under Grant 90303024 and the
Guangdong Province Natural Science Foundation under Grant 020146 and Grant
031541 of China.}}
\author{
Daowen Qiu\thanks{ D. Qiu is with Department of Computer Science,
Zhongshan University, Guangzhou, 510275, China (e-mail:
issqdw@mail.sysu.edu.cn).}}
\date{  }
\maketitle
\date{  }
\maketitle
\begin{center}
\begin{minipage}{150mm}

{\bf Abstract:} Fuzzy discrete event systems (DESs) were proposed
recently by Lin and Ying [19], which may better cope with the
real-world problems with fuzziness, impreciseness, and
subjectivity such as those in biomedicine. As a continuation of
[19], in this paper we further develop fuzzy DESs by dealing with
supervisory control of fuzzy DESs. More specifically, (i) we
reformulate the parallel composition of crisp DESs, and then
define the parallel composition of fuzzy DESs that is equivalent
to that in [19];  {\it max-product} and {\it max-min} automata for
modeling fuzzy DESs are considered; (ii) we deal with a number of
fundamental problems regarding supervisory control of fuzzy DESs,
particularly demonstrate controllability theorem and nonblocking
controllability theorem of fuzzy DESs, and thus present the
conditions for the existence of supervisors in fuzzy DESs; (iii)
we analyze the complexity for presenting a uniform criterion to
test the fuzzy controllability condition of fuzzy DESs modeled by
max-product automata; in particular, we present in detail a
general computing method for checking whether or not the fuzzy
controllability condition holds, if max-min automata are used to
model fuzzy DESs, and by means of this method we can search for
all possible fuzzy states reachable from initial fuzzy state in
max-min automata;  also, we introduce the fuzzy
$n$-controllability condition for some practical problems; (iv) a
number of examples serving to illustrate the applications of the
derived results and methods are described; some basic properties
related to supervisory control of fuzzy DESs are investigated. To
conclude, some related issues are raised for further
consideration.

\par
\hskip 5mm


\end{minipage}
\end{center}
\smallskip

\section*{I. Introduction}

A Discrete Event System (DES) is a discrete-state, event-driven
dynamical system in which the state space is described by a
discrete set, and states evolve in terms of asynchronous
occurrence of discrete events over time [11,1,26,27]. Actually,
DESs represent many technological and engineering systems such as
communication networks, computer networks, manufacturing systems,
and transportation systems. So far, many modeling approaches of
DESs have been proposed and developed, including the notably
finite automata [26,11,1,31], Petri nets [22,7,1], and their
variations [1].

In DESs, an important research direction is the supervisory
control, which was launched by Ramadge and Wonham [26,27] in the
1980's. Since then, many researchers have also made contributions
to the developments and applications of DES theory (for example,
[11,1,31] and the references therein). In the study of supervisory
control, a DES can be well modeled by a formal language of finite
strings of event symbols, and the set of all possible events is
partitioned into both controllable and uncontrollable events. In
this framework, a supervisory mechanism is established, in which
the supervisor may enable or disable the controllable events in
order to satisfy a set of qualitative specifications on the
admissible occurrences of the events that can be executed by the
system. Moreover, the supervisor may not be able to observe all of
the events generated by the system due to the existence of
unobservable events in the set of events.

In order to understand more intuitively supervisory control of
crisp DESs under complete observation (all events can be seen),
let us  further introduce it more formally. (For the details, we
refer to [1,11].)  Here we assume that the given DES is modeled by
a finite automaton $G$. The behavior of the DES is characterized
by the language $L(G)$ generated by $G$, but the behavior is not
satisfactory, that is, $L(G)$ contains strings that are not
allowed. Thus, our objective is to consider the sublanguages of
$L(G)$ that stand for the legal or admissible behavior for the
controlled system. For example, if $K\subseteq L(G)$ is a legal
sublanguage of $L(G)$, then we attempt to design a supervisor,
denoted by $S$, such that $S$ interacting $G$ will result in the
behavior desired. In general, $S$ has the capability of disabling
some feasible events of $G$, and the set of feasible events is
termed as controllable events. If $S$ exists, then $K$ is called
to be controllable; otherwise, we hope to find the largest
controllable sublanguage and the smallest controllable
superlanguage of $K$. Indeed, computing for finding these
sublanguages is related to classical formal language and automata
theory [10,1,11].

Though most of the research on supervisory control of DESs has
focused on systems that are modeled as deterministic finite
automata, there are also interesting studies concerning
nondeterministic DESs [8] and probabilistic DESs (see [16] and the
references therein) for dealing with some uncertainty problems and
complex information and systems in practice. However, in real-life
situation, there are also a large number of problems with
vagueness, impreciseness, and subjectivity that can be reasonably
described by the use of words instead of numbers, where words mean
the possibility distributions suggested by Zadeh [38], that is,
fuzzy subsets. For example, ``young", ``old", ``small", ``large",
``excellent", ``good", and ``poor", etc. can be viewed as the
words that are possibility distributions. Actually, computing with
words, as a methodology, was advocated by Zadeh [38,39] and has
been significantly developed by others [40], which may conform
more to human's perception when describing the real-world
problems. Therefore, recently, Lin and Ying [18,17] initiated
significantly the study of fuzzy DESs by combining fuzzy set
theory with DESs, with a desire to solve those problems not
capable of being satisfactorily treated by conventional DESs.

Exactly, there are a large number of DESs to be treated better in
a manner of computing with words when describing their states.
This is especially true in biomedical applications, as stressed in
[19]. For example, if we describe the health condition of a person
with ``excellent" (E), ``fair" (F), and ``poor" (P), then each
state has certain possibility, and we thus write the condition
with a word ``health condition" ($HC$):
\[
HC=a/E+b/F+c/P
\]
where $a,b,c\in[0,1]$ represent the possibilities of E, F, and P,
respectively. Equivalently, as in [19], the above word can also be
represented as a vector $[a \hskip 2mm b \hskip 2mm c]$. In
reality, this representation leads to the formulation of fuzzy
finite automata and fuzzy DESs [19,18]. It is worth indicating
that fuzzy finite automata were discussed long ago by Wee [34],
Santos [28,29], Lee and Zadeh [20], as well as others
[32,9,12,3,21,23,24,25] (the more references are referred to [21]
therein). In practice, fuzzy finite automata and fuzzy languages
have been used to solve meaningful problems such as intelligent
interface design [5], clinical monitoring [30], neural networks
[33,4],  and pattern recognition by DePalma and  Yau (see [3,21]
for the details). As well, fuzzy finite automata can be viewed as
a type of formal models for computing with words [37,36] when the
inputs are strings of words rather than symbols. However, as the
author is aware, fuzzy DESs modeled by fuzzy finite automata were
not taken into account until the work of Refs. [19,18] by Lin and
Ying.

In [19] the authors formalized a fuzzy DES as a fuzzy finite
automaton. More exactly, by means of [19] a fuzzy state is
represented as a form of vector that can be viewed as a
possibility distribution over the set of all considered states,
that is, a word, and a fuzzy event $\tilde{\sigma}$ is
equivalently represented as a matrix in which each element, say
$a_{ij}$, belongs to [0,1], meaning the possibility for the
current state $q_{i}$ changing to state $q_{j}$ when event
$\sigma$ occurs. As well, the parallel composition of fuzzy finite
automata was discussed in [19]. With the appropriate formulation,
Lin and Ying [19] dealt with observability and state estimation of
fuzzy DESs. Furthermore, they investigated some optimal control
problems. Thus, a framework of fuzzy DES theory has been primarily
established in [19].

In supervisory control theory, how to deal with the existence of
supervisor in the presence of uncontrollable events is a very
fundamental issue, which was described in [1] with a number of key
theorems, including Controllability Theorem and Nonblocking
Controllability Theorem, and Controllability and Observability
Theorem. However, these theorems in the setting of fuzzy DESs
still have not been discussed. As pointed out in [19, page 409,
lines 6-10], a comprehensive theory of fuzzy DESs still needs to
be set up, including many fundamentally important concepts,
methods, and theorems in the traditional DESs, such as
controllability, observability, and optimal control. Hence, as a
continuation of [19], the principal goal of this paper is to deal
with controllability theorems of fuzzy DESs and some related
problems.

More specifically, our main aim in the present paper is to
consider a number of fundamental issues concerning supervisory
control of fuzzy DESs, and the related properties. The rest of the
paper is organized as follows. In Section II we review the
formulation of fuzzy DESs, and define fuzzy DESs modeled by {\it
max-product} and {\it max-min} automata; as well, we reformulate
the parallel composition of crisp DESs, with which a parallel
composition of fuzzy DESs equivalent to that in [19] is defined;
notably, the computing processes in the newly defined operation
may be more succinct and convenient. The focus of Section III is
to deal with a number of fundamental problems of supervisory
control of fuzzy DESs; we demonstrate controllability theorem and
nonblocking controllability theorem of fuzzy DESs, and thus
present the conditions for the existence of supervisors in fuzzy
DESs; we analyze the complexity for presenting a uniform criterion
to check the fuzzy controllability condition of fuzzy DESs modeled
by {\it max-product} automata; in particular, if fuzzy DESs are
modeled by {\it max-min} automata, then we present in detail a
general method for checking whether or not the fuzzy
controllability condition holds, and by means of this method we
can search for all possible fuzzy states reachable from initial
fuzzy state in max-min automata; furthermore, the fuzzy
$n$-controllability condition is introduced for some practical
problems; a number of examples are described to serve as
illustrating the use of the given methods and results. Section IV
investigates the properties related to supervisory control of
fuzzy DESs, including the existence of supremal fuzzy controllable
sublanguage and infimal prefix-closed fuzzy controllable
superlanguage of a given fuzzy noncontrollable sublanguage.
Finally,  in Section V we summarize the main results obtained, and
raise a number of related issues for further consideration.

Concepts and notations used in this paper will be given and
explained when they first appear.

\section*{II. Fuzzy Discrete Event Systems}

\subsection*{{\it A. Language and Automaton Models of DESs}}

In this subsection, we briefly review some basic concepts from
formal languages and automata theory [10].  Let $\Sigma$ denote
the set of events of a DES, and let $\Sigma^{*}$ denote the set of
all finite length strings of events from $\Sigma$, including empty
string denoted by $\epsilon$. Each sequence of events is called a
{\it string} of the system, and a collection of strings is called
a {\it language}. A language is {\it regular} if it is {\it
marked} (or {\it recognized}) by a finite automaton, which is
defined by what follows. A nondeterministic finite automaton,
denoted by $G$, is a system described by $
G=(Q,\Sigma,\delta,q_{0},Q_{m})$, where $Q$ is the finite set of
states; $\Sigma$ is the finite set of events; $\delta: Q\times
\Sigma\rightarrow {\cal P}(Q)$ is the transition function (In what
follows, ${\cal P}(X)$ represents the power set of set $X$);
$q_{0}\in Q$ is the initial state; and $Q_{m}\subseteq Q$ is
called the set of marked states. Indeed, transition function
$\delta$ can be naturally extended to $Q\times\Sigma^{*}$ in the
following manner: For any $q\in Q$, any $s\in\Sigma^{*}$ and
$\sigma\in\Sigma$, $ \delta(q,\epsilon)=q,\hskip 2mm {\rm and}
\hskip 2mm \delta(q,s\sigma)=\delta(\delta(q,s),\sigma), $ where
notably we define $\delta(A,\sigma)=\bigcup_{q\in
A}\delta(q,\sigma)$ for any $A\in {\cal P}(Q)$. As in [17], we
represent equivalently states and events by the forms of vectors
and matrices, respectively. More specifically, for finite
automaton $G=(Q,\Sigma,\delta,q_{0},Q_{m})$, where
$Q=\{q_{1},q_{2},\ldots,q_{n}\}$, we represent $q_{i}$ by vector
$\bar{s}_{i}=[0\hskip 2mm\cdots\hskip 2mm 1\hskip 2mm \cdots\hskip
2mm 0]$ where 1 is in the $i$th place and the dimension equals
$n$; for $\sigma\in\Sigma$, $\sigma$ is represented as a 0-1
matrix $[a_{ij}]_{n\times n}$ where $a_{ij}\in \{0,1\}$, and
$a_{ij}=1$ if and only if $q_{j}\in\delta (q_{i},\sigma)$.
Analogously, vector $[0\hskip 2mm\cdots\hskip 2mm 1\hskip 2mm
0\hskip 2mm \cdots\hskip 2mm 1\hskip 2mm\cdots\hskip 2mm 0]$ in
which 1 is in the $i$th and $j$th places, respectively, means that
the current state may be $q_{i}$ or $q_{j}$.

In order to define and better understand  parallel composition of
fuzzy finite automata, we reformulate the parallel composition of
crisp finite automata [1,10,11,19]. For finite automata
$G_{i}=(Q_{i},\Sigma_{i},\delta_{i},q_{0i},Q_{mi})$, $i=1,2$, we
reformulate the parallel composition in terms of the following
fashion:
\[
G_{1}\|^{'}G_{2}=(Q_{1}\otimes Q_{2}, \Sigma_{1}\cup\Sigma_{2},
\delta_{1}\|^{'} \delta_{2},q_{10}\otimes q_{20},Q_{m1}\otimes
Q_{m2}).
\]
Here, $Q_{1}\otimes Q_{2}=\{q_{1}\otimes q_{2}:q_{1}\in
Q_{1},q_{2}\in Q_{2}\}$, and symbol $``\otimes"$ denotes tensor
product that is different from the composition of fuzzy relation
that means max-min operation [19]. For event $\sigma\in
\Sigma_{1}\cup\Sigma_{2}$, we define the corresponding matrix of
$\sigma$ in $G_{1}\|^{'} G_{2}$ as follows:

(i) If event $\sigma\in \Sigma_{1}\cap \Sigma_{2}$, then
$\sigma=\sigma_{1}\otimes\sigma_{2}$ where $\sigma_{1}$ and
$\sigma_{2}$ are the matrices of $\sigma$ in $G_{1}$ and $G_{2}$,
respectively.

(ii) If event $\sigma\in\Sigma_{1}\backslash\Sigma_{2}$, then
$\sigma=\sigma_{1}\otimes I_{2}$ where $\sigma_{1}$ is the matrix
of $\sigma$ in $G_{1}$, and $I_{2}$ is unit matrix of order
$|Q_{2}|$. (In this paper, $|X|$ denotes the number of all
elements in set $X$.)

(iii) If event $\sigma\in\Sigma_{2}\backslash\Sigma_{1}$, then
$\sigma= I_{1}\otimes\sigma_{2}$ where $\sigma_{2}$ is the matrix
of $\sigma$ in $G_{2}$ and $I_{1}$ is unit matrix of order
$|Q_{1}|$.

In terms of the above (i-iii) regarding the event $\sigma\in
\Sigma_{1}\cup\Sigma_{2}$, we can define $\delta_{1}\|^{'}
\delta_{2}$ as: For $ q_{1}\otimes q_{2} \in Q_{1}\otimes Q_{2}$,
$\sigma\in \Sigma_{1}\cup\Sigma_{2}$,
\[
(\delta_{1}\|^{'} \delta_{2})(q_{1}\otimes q_{2},\sigma)=\left\{
\begin{array}{lll}
(q_{1}\otimes q_{2} )\times(\sigma_{1}\otimes\sigma_{2}),&{\rm if}& \sigma\in \Sigma_{1}\cap \Sigma_{2},\\
(q_{1}\otimes q_{2} )\times(\sigma_{1}\otimes I_{2}),&{\rm if}& \sigma\in \Sigma_{1} \backslash \Sigma_{2},\\
(q_{1}\otimes q_{2} )\times (I_{1}\otimes\sigma_{2}),&{\rm if}& \sigma\in \Sigma_{2} \backslash \Sigma_{1},
\end{array}
\right.
\]
where $\times$ is the usual product between matrices, and, as
indicated above, symbol $\otimes$ denotes tensor product of
matrices, that is, for $m\times n$ matrix $A= [a_{ij}]_{m\times
n}$ and $k\times l$ matrix $B= [b_{ij}]_{k\times l}$, we have
\[
A\otimes B=\left[\begin{array}{lll}
a_{11}B&\cdots & a_{1n}B\\
\vdots &\ddots & \vdots\\
a_{m1}B &\cdots & a_{mn}B
\end{array}
\right].
\]

{\it Remark 1.} One can check that for any finite automata $G_{1}$
and $G_{2}$ with event sets $\Sigma_{1}$ and $\Sigma_{2}$,
respectively, the  parallel composition defined above is
equivalent to the usual one, that is,
\[
L( G_{1}\| G_{2} )=L(G_{1}\|^{'}G_{2}) \hskip 2mm {\rm and} \hskip
2mm L_{m}(G_{1}\| G_{2})=L_{m}(G_{1}\|^{'} G_{2}),
\]
where $G_{1}\| G_{2} $ is the usual parallel composition. In
particular, when $\Sigma_{1}=\Sigma_{2}=\Sigma$, the parallel
composition $\|^{'}$ reformulated above degenerates to the product
operation [1,10]. We may see that defining parallel composition of
finite automata by means of tensor product is more suitable to
describe the composition operation with fuzzy states of each which
may simultaneously belong to several crisp states with respective
memberships, and with fuzzy events of each which may result in a
transition between any two states with certain degree.

\subsection*{{\it B.  Language and Automaton Models of Fuzzy DESs}}

In light of Subsection A, in the setting of fuzzy DESs, a fuzzy
state is naturally represented as a vector $[a_{1}\hskip 2mm
a_{2}\hskip 2mm\cdots\hskip 2mm a_{n}]$ where $a_{i}\in[0,1]$
represents the possibility of the current state being $q_{i}$, and
$n$ stands for the number of all possible crisp states. Similarly,
a fuzzy event is denoted by a matrices $[a_{ij}]_{n\times n}$ as
well, but here elements $a_{ij}\in [0,1]$ rather than $\{0,1\}$,
and $a_{ij}$ means the possibility for the system to transit from
the current state $q_{i}$ to state $q_{j}$ when event $\sigma$
occurs. Hence, a fuzzy finite automaton is defined as a fuzzy
system
$\tilde{G}=(\tilde{Q},\tilde{\Sigma},\tilde{\delta},\tilde{q}_{0},\tilde{Q}_{m})$,
where $\tilde{Q}$ is a set of some possibility distributions
(fuzzy states) over crisp state set $Q$; $\tilde{q}_{0}$ is the
initial fuzzy state; $\tilde{Q}_{m}$ is also a set of fuzzy states
over $Q$, standing for the marking states; $\tilde{\Sigma}$ is a
set of fuzzy events (matrices);
$\tilde{\delta}:\tilde{Q}\times\tilde{\Sigma}\rightarrow\tilde{Q}$
is a transition function which is defined by
$\tilde{\delta}(\tilde{q},\tilde{\sigma})= \tilde{q}\odot
\tilde{\sigma}$ or $\tilde{q}\circ \tilde{\sigma}$ for
$\tilde{q}\in \tilde{Q}$ and $\tilde{\sigma}\in\tilde{\Sigma}$,
where $\odot$ and $\circ$ denote respectively {\it max-min} and
{\it max-product} operations in fuzzy set theory [13]: For
$n\times m$ matrix $A=[a_{ij}]$ and $m\times k$ matrix
$B=[b_{ij}]$, then the elements in $n\times k$ matrixes $A\odot
B=C_{1}$ and $A\circ B=C_{2}$ are respectively as
$c_{ij}^{(1)}=\max_{l=1}^{m} \min\{a_{il}, b_{lj}\}$ and
$c_{ij}^{(2)}=\max_{l=1}^{m} a_{il}\times b_{lj}$. As a complete
system, we usually require that $ \tilde{q}_{0} \in\tilde{Q}$ and
$\tilde{Q}_{m}\subseteq \tilde{Q}$. In the sequel, a fuzzy finite
automaton defined in terms of the  max-min (resp.  max-product)
operation for its transition function is said to be {\it max-min}
(resp. {\it max-product}) automaton.

The fuzzy languages generated and marked by $\tilde{G}$, denoted
by ${\cal L}_{\tilde{G}}$ and ${\cal L}_{\tilde{G},m}$,
respectively, are defined as a function from $\tilde{\Sigma}^{*}$
($\tilde{\Sigma}^{*}$ represents the set of all strings of fuzzy
events from $\tilde{\Sigma}$) to [0,1] as follows: For any $
\tilde{\sigma}_{1}\tilde{\sigma}_{2}\cdots\tilde{\sigma}_{k} \in
\tilde{\Sigma}^{*}$ where $\tilde{\sigma}_{i}\in\tilde{\Sigma}$,
$i=1,2,\cdots,k$,
\begin{equation}
{\cal L}_{\tilde{G}}
(\tilde{\sigma}_{1}\tilde{\sigma}_{2}\cdots\tilde{\sigma}_{k})=\max_{i=1}^{n}\tilde{q}_{0}*
\tilde{\sigma}_{1}*\tilde{\sigma}_{2}*\cdots*\tilde{\sigma}_{k}*
\bar{s}_{i}^{T},
\end{equation}
\begin{equation}
{\cal
L}_{\tilde{G},m}(\tilde{\sigma}_{1}\tilde{\sigma}_{2}\cdots\tilde{\sigma}_{k})=\sup_{\tilde{q}\in
\tilde{Q}_{m}} \tilde{q}_{0}*
\tilde{\sigma}_{1}*\tilde{\sigma}_{2}*\cdots*\tilde{\sigma}_{k}*
\tilde{q}^{T},
\end{equation}
where  $*$ is either max-min or max-product operation in terms of
the type of  $\tilde{G}$;  $\bar{s}_{i}^{T}$ is the transpose of
$\bar{s}_{i}$, and $\bar{s}_{i}$ is as indicated above, i.e.,
$\bar{s}_{i}=[0\hskip 2mm \cdots\hskip 2mm 1 \hskip 2mm
\cdots\hskip 2mm 0]$ where 1 is in the $i$th place.

Intuitively, ${\cal L}_{\tilde{G}} (
\tilde{\sigma}_{1}\tilde{\sigma}_{2}\cdots\tilde{\sigma}_{k} )$
represents the degree of the string of fuzzy events
$\tilde{\sigma}_{1}\tilde{\sigma}_{2}\cdots\tilde{\sigma}_{k}$
being physically possible, while ${\cal
L}_{\tilde{G},m}(\tilde{\sigma}_{1}\tilde{\sigma}_{2}\cdots\tilde{\sigma}_{k})$
stands for the possibility for the same string being marked
(recognized) by the fuzzy automaton $\tilde{G}$.  Clearly, from
Eqs. (1) and (2) it follows that for any $ \tilde{ s} \in
\tilde{\Sigma}^{*}$ and any $ \tilde{\sigma} \in \tilde{\Sigma}$,
\begin{equation}
{\cal L}_{\tilde{G},m}( \tilde{ s}  \tilde{\sigma}  )\leq
{\cal L}_{\tilde{G}}( \tilde{ s} \tilde{\sigma} )\leq
{\cal L}_{\tilde{G}}( \tilde{ s} )
\end{equation}
which means that the degree of a string of fuzzy events being
physically possible is not smaller than that of it being marked
(the first inequality) and is not bigger than that of its any
substring being physically possible (the second inequality).

{\it Remark 2.} The max-product automata are equivalent to the
fuzzy automata defined by Lin and Ying [19], and in any
max-product automaton with initial state $\tilde{q}_{0}$ and input
alphabet $\tilde{\Sigma}$, the set of states $\{\tilde{q}_{0}\circ
\tilde{s}:\tilde{s}\in \tilde{\Sigma}^{*}\}$ is likely infinite,
and, as we will see in the sequel, it may result in complicated
computation in realization. Therefore, we would like to consider
max-min automata usually in practical applications. In reality,
max-min automata are similar to the fuzzy automata defined in [30]
for dealing with an application of clinical monitoring. Notably,
the set of fuzzy states $\{\tilde{q}_{0}\odot
\tilde{s}:\tilde{s}\in \tilde{\Sigma}^{*}\}$ in any max-min
automaton is clearly finite [30].

We introduce the parallel operation of fuzzy finite automata
(max-product or max-min automata). For fuzzy finite automata
$\tilde{G}_{i}=(\tilde{Q}_{i},\tilde{\Sigma}_{i},\tilde{\delta}_{i},\tilde{q}_{i0},\tilde{Q}_{mi})$,
$i=1,2$, by means of the crisp case formulated above we can
naturally define the parallel composition of $\tilde{G}_{1}$ and
$\tilde{G}_{2}$:
\[
\tilde{G}_{1}\|^{'} \tilde{G}_{2} =( \tilde{Q}_{1}\otimes
\tilde{Q}_{2}, \tilde{\Sigma}_{1}\cup\tilde{\Sigma}_{2},
\tilde{\delta}_{1}\|^{'} \tilde{\delta}_{2}, \tilde{q}_{10}\otimes
\tilde{q}_{20},\tilde{Q}_{m1}\otimes \tilde{Q}_{m2})
\]
where $\tilde{Q}_{1}\otimes \tilde{Q}_{2}=\{\tilde{q}_{1}\otimes
\tilde{q}_{2}:\tilde{q}_{i},i=1,2\}$,
$\tilde{\Sigma}_{1}\cup\tilde{\Sigma}_{2}=\{\tilde{\sigma}:\sigma\in\Sigma_{1}\cup\Sigma_{2}\}$
where $\tilde{\sigma}$ is defined as follows:

(i) If $\sigma\in \Sigma_{1}\cap\Sigma_{2}$, then
$\tilde{\sigma}=\tilde{\sigma}_{1}\otimes\tilde{\sigma}_{2}$ where
$\tilde{\sigma}_{i}$ is the matrix of event $\sigma$ in
$\tilde{G}_{i}$, $i=1,2$.

(ii) If event $\sigma\in\Sigma_{1} \backslash\Sigma_{2}$, then
$\tilde{\sigma}=\tilde{\sigma}_{1}\otimes I_{2}$ where
$\tilde{\sigma}_{1}$  is the matrix of event $\sigma$ in
$\tilde{G}_{1}$ and $I_{2}$ is unit matrix of order  $|Q_{2}|$.

(iii) If event $\sigma\in\Sigma_{2}\backslash\Sigma_{1}$, then
$\tilde{\sigma}= I_{1}\otimes\tilde{\sigma}_{2}$ where $I_{1}$ is
unit matrix of order $|Q_{1}|$ and $\tilde{\sigma}_{2}$ is equal
to the matrix of $\sigma$ in $\tilde{G}_{2}$.

For any $ \tilde{q}_{1}\otimes \tilde{q}_{2}\in
\tilde{Q}_{1}\otimes \tilde{Q}_{2}$ and any $\tilde{\sigma}\in
\tilde{\Sigma}_{1}\cup\tilde{\Sigma}_{2}$,
\[
(\tilde{\delta}_{1}\|^{'} \tilde{\delta}_{2})(\tilde{q}_{1}\otimes
\tilde{q}_{2},\tilde{\sigma})= (\tilde{q}_{1}\otimes
\tilde{q}_{2})\circ\tilde{\sigma}.
\]

{\it Remark 3.} The parallel composition $\tilde{G}_{1}\|^{'}
\tilde{G}_{2}$ of max-product automata formulated above is
equivalent to that in [19]. In [19] the parallel composition of
fuzzy automata $\tilde{G}_{1}$ and $\tilde{G}_{2}$ is denoted by
$\tilde{G}_{1}\|\tilde{G}_{2}=(\tilde{Q}_{1}\times\tilde{Q}_{2},
\tilde{\Sigma}_{1}\cup\tilde{\Sigma}_{2},\tilde{\delta}_{1}\times\tilde{\delta}_{2},(\tilde{q}_{10},
\tilde{q}_{20}))$ where every state in
$\tilde{Q}_{1}\times\tilde{Q}_{2}$ is represented as the form of
Cartesian product $(\tilde{q}_{1},\tilde{q}_{2})$ for
$\tilde{q}_{1}\in\tilde{Q}_{1}$and
$\tilde{q}_{2}\in\tilde{Q}_{2}$. And it should be pointed out that
the authors [19, page 411] also considered implicitly tensor
product for the computing process but it was not formulated to
establish the parallel composition by means of tensor product. To
a certain extent, it may be more clear and convenient to describe
the computing processes in terms of the parallel composition
defined in this paper. An example in the following is provided to
illustrate the computing.

{\it Example 1.}  In fuzzy finite automata $\tilde{G}_{1}$ and
$\tilde{G}_{2}$ defined above, if $\tilde{p}=[0.1\hskip 2mm
0.5\hskip 2mm 0.3]\in\tilde{Q}_{1}$, $\tilde{q}=[0.2\hskip 2mm
0.6\hskip 2mm 0.1]\in\tilde{Q}_{2}$,
$\tilde{\alpha}_{1}=\left[\begin{array}{lcc} 0.1&0.2&0\\
0.4&0&0.7\\
0.6&0.8&0
\end{array}
\right]\in \tilde{\Sigma}_{1}$ (here $\alpha_{1}\in \Sigma_{1}\backslash\Sigma_{2}$), then

$
\tilde{p}\otimes\tilde{q}=[0.02\hskip 2mm 0.06\hskip 2mm
0.01\hskip 2mm 0.1\hskip 2mm 0.3\hskip 2mm 0.05\hskip 2mm
0.06\hskip 2mm 0.18\hskip 2mm 0.03],
$

$ \tilde{\alpha}_{1}\otimes I_{2}=\tilde{\alpha}_{1}\otimes \left[\begin{array}{lll} 1&0&0\\
0&1&0\\
0&0&1
\end{array}
\right] =\left[\begin{array}{lllllllll} 0.1&0&0&0.2&0&0&0&0&0\\
0&0.1&0&0&0.2&0&0&0&0\\
0&0&0.1&0&0&0.2&0&0&0\\
0.4&0&0&0&0&0&0.7&0&0\\
0&0.4&0&0&0&0&0&0.7&0\\
0&0&0.4&0&0&0&0&0&0.7\\
0.6&0&0&0.8&0&0&0&0&0\\
0&0.6&0&0&0.8&0&0&0&0\\
0&0&0.6&0&0&0.8&0&0&0
\end{array}
\right],
$\\
and therefore, if $\tilde{G}_{1}$ and $\tilde{G}_{2}$ are max-min
automata, then
\begin{eqnarray*}
\tilde{\delta}_{1}\|^{'}\tilde{\delta}_{2}(\tilde{p}\otimes\tilde{q},
\tilde{\alpha}_{1} )&=&
(\tilde{p}\otimes\tilde{q})\odot ( \tilde{\alpha}_{1}\otimes I_{2} ) \\
&=&[0.1\hskip 2mm 0.3\hskip 2mm 0.05\hskip 2mm 0.06\hskip 2mm
0.18\hskip 2mm 0.03\hskip 2mm 0.1\hskip 2mm 0.3\hskip 2mm 0.05];
\end{eqnarray*}
if $\tilde{G}_{1}$ and $\tilde{G}_{2}$ are max-product automata,
then
\begin{eqnarray*}
\tilde{\delta}_{1}\|^{'}\tilde{\delta}_{2}(\tilde{p}\otimes\tilde{q}, \tilde{\alpha}_{1} )&=&
(\tilde{p}\otimes\tilde{q})\circ ( \tilde{\alpha}_{1}\otimes I_{2} ) \\
&=&[\max\{0.002,0.004,0.036\}\hskip 2mm \max\{0.006,0.12,0.108\}\hskip 2mm \max\{0.001,0.02,0.18\}\\
&&\max\{0.004,0.48\}\hskip 2mm\max\{0.012,0.144\}\hskip 2mm\max\{0.002,0.24\}\hskip 2mm 0.07 \hskip 2mm 0.21 \hskip 2mm 0.35]\\
&=&[0.036\hskip 2mm 0.12\hskip 2mm 0.18\hskip 2mm 0.48\hskip 2mm
0.144\hskip 2mm 0.24\hskip 2mm 0.07\hskip 2mm 0.21\hskip 2mm
0.35].\hskip 5mm \Box
\end{eqnarray*}

\section*{III. Supervisory Controllability of Fuzzy DESs}

\subsection*{{\it A. Controllability Theorems for Fuzzy DESs}}

In supervisory control, a fundamental issue is how to design a
controller (or supervisor) whose task is to enable and disable the
controllable events such that the resulting closed-loop system
obeys some pre-specified operating rules [1]. The purpose of this
subsection is to establish controllability theorem of fuzzy DESs.
Again, let us recall the supervisory control problems of crisp
DESs under full observation. Suppose that a DES is modeled by a
finite automaton $G=(Q,\Sigma,\delta,q_{0},Q_{m})$ and the
language $L(G)$ generated by $G$ may be interpreted as physically
possible behavior. However, $L(G)$ is not satisfactory and must be
modified by feedback control; modifying the behavior is to be
understood as restricting the behavior to a subset of $L(G)$ that
represents the ``legal"  behavior for the controlled system. In
order to alter the behavior of $L(G)$, a supervisor denoted by
$S$, is introduced. As a supervisor, $S$ has the capability of
disabling some, but not necessarily all events of $G$. In this
regard, $\Sigma$ is partitioned into two disjoint subsets
$\Sigma=\Sigma_{c}\cup\Sigma_{uc}$ where $\Sigma_{c}$ is the set
of controllable events that can be prevented from happening or
disabled, by supervisor $S$, whereas $\Sigma_{uc}$ is the set of
uncontrollable events which cannot be prevented from happening by
$S$. Formally, supervisor $S$ is defined as a function:
\[
S: L(G)\rightarrow {\cal P}(\Sigma).
\]
It is interpreted that for each $s\in L(G)$, $S(s)\cap
\{\sigma:s\sigma\in L(G)\}$ represents the set of enabled events
after the occurrence of $s$. Furthermore, it is required that for
any $s\in L(G)$,
\begin{equation}
\Sigma_{uc}\cap \{\sigma\in \Sigma:s\sigma\in L(G)\}\subseteq S(s),
\end{equation}
which means that after the occurrence of any physically possible
string of events, the physically possible  uncontrollable events
are not allowed to be disabled by $S$. The condition described by
Eq. (4) is called {\it admissible} for $S$ [1].

Given a DES modeled by finite automaton $G$ and admissible
supervisor $S$, the resulting controlled system denoted by $S/G$
that means $S$ controlling $G$, is also a DES modeled by a
language $L(S/G)$ defined recursively as follows: $\epsilon \in
L(S/G)$ and

$s\sigma\in L(S/G)\Longleftrightarrow s\in L(S/G)$ and $s\sigma \in L(G)$ and $\sigma\in S(s)$.\\
The language marked by $S/G$ is defined as:
\[
L_{m}(S/G)=L(S/G)\cap L_{m}(G).
\]
The DES modeled by $L(S/G)$ is {\it blocking} if $L(S/G)\not= pr
(L_{m}(S/G)).$ As mentioned before, a key result in supervisory
control is the controllability theorem that characterizes the
existence for supervisors in the presence of uncontrollable
events, and is described as follows: Suppose that a DES is modeled
by finite automaton $G=(Q,\Sigma,\delta,q_{0})$. (Here the marked
set is not considered and thus left out.)  Let $K\subseteq L(G)$.
Then there is supervisor $S$ such that $L(S/G)=pr(K)$ if and only
if $pr(K)\Sigma_{uc}\cap L(G)\subseteq pr(K)$.

As stated before, imprecision and uncertainty play a crucial role
in the field of biomedicine. For example, in practice, a single
disease may show several different symptoms with respective
severity degrees at every therapeutic stage. With the progress of
treatment, some bad symptoms may fade away while some healthy
conditions appear; also, some symptoms may weaken and the other
ones possibly strength. We formally denote by $p_{1},p_{2},\ldots,
p_{n}$ the possible main symptoms, containing healthy conditions
and poor ones. If the current condition of a patient is with
respective grades of membership $d_{i}\in [0,1]$ for $p_{i}$,
denoted by $\sum_{i}d_{i}^{(1)}/p_{i}$, then a therapy (say
$\tilde{\alpha}$) may result in another condition
$\sum_{i}d_{i}^{(2)}/p_{i}$.  Therefore, we will try to model such
a process by means of fuzzy DESs. Furthermore, in view of  human
errors in medicine occurring frequently in hospitalized patients
[14], we focus on dealing with supervisory control of fuzzy DESs
to try to prevent medical errors such as those in diagnosis and
treatment.  (Fig. 1 and Fig. 2 manifest respectively the rough
processes of treat without supervisor and under supervisor.)

\setlength{\unitlength}{0.05in}

\begin{picture}(50,50)

\put(0,30){\line(1,0){10}} \put(10,27){\line(0,1){6}}

\put(10,27){\line(1,0){15}}\put(25,27){\line(0,1){6}}

\put(25,33){\line(-1,0){15}}\put(35,30){\vector(-1,0){10}}

\put(0,30){\line(0,-1){15}}\put(0,15){\vector(1,0){10}}
\put(0,15){\line(1,0){10}}\put(10,18){\line(0,-1){6}}
\put(10,18){\line(1,0){15}}\put(25,12){\line(0,1){6}}
\put(25,12){\line(-1,0){15}}\put(25,15){\line(1,0){10}}
\put(35,15){\line(0,1){15}}

\put(10,27){\makebox(15,6)[c]{{\tiny Patient}}}
\put(10,12){\makebox(15,6)[c]{{\tiny Physician}}}
\put(-3,18.5){\makebox(15,6)[c]{{\tiny Feedback}}}
\put(23,18.5){\makebox(15,6)[c]{{\tiny Treat}}}
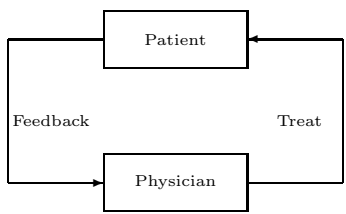
\put(10,3){\makebox(15,6)[c]{{\tiny Fig.  1. Treat without
supervisor.}}}

\put(55,45){\line(1,0){10}} \put(65,42){\line(0,1){6}}

\put(65,42){\line(1,0){15}}\put(80,42){\line(0,1){6}}

\put(80,48){\line(-1,0){15}}\put(95,45){\vector(-1,0){15}}

\put(55,45){\line(0,-1){30}}\put(55,30){\vector(1,0){10}}
\put(55,30){\line(1,0){10}}\put(65,33){\line(0,-1){6}}
\put(65,33){\line(1,0){15}}\put(80,27){\line(0,1){6}}
\put(80,27){\line(-1,0){15}}\put(80,15){\line(1,0){15}}
\put(95,15){\line(0,1){30}}

\put(55,15){\vector(1,0){10}}\put(65,18){\line(0,-1){6}}
\put(65,18){\line(1,0){15}}\put(80,12){\line(0,1){6}}
\put(80,12){\line(-1,0){15}}\put(72.5,27){\vector(0,-1){9}}

\put(65,42){\makebox(15,6)[c]{{\tiny Patient}}}
\put(65,27){\makebox(15,6)[c]{{\tiny Physician}}}
\put(53,33.5){\makebox(15,6)[c]{{\tiny Feedback}}}
\put(75,33.5){\makebox(15,6)[c]{{\tiny Supervised Treat Method}}}
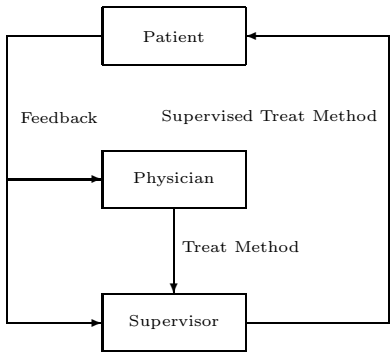
\put(65,3){\makebox(15,6)[c]{{\tiny Fig.  2. Treat under
supervisor.}}} \put(65,12){\makebox(15,6)[c]{{\tiny Supervisor}}}
\put(72,20){\makebox(15,6)[c]{{\tiny Treat Method}}}

\end{picture}

\vskip -5mm

Before giving controllability theorem in fuzzy DESs, we need
further to define several related concepts in the setting of fuzzy
DESs. Each event $\tilde{\sigma}\in\tilde{\Sigma}$  is associated
with a degree of controllability, so, the uncontrollable set
$\tilde{\Sigma}_{uc}$ and controllable set $\tilde{\Sigma}_{c}$
are two fuzzy subsets of $\tilde{\Sigma}$, i.e.,
$\tilde{\Sigma}_{uc}, \tilde{\Sigma}_{c}\in {\cal
F}(\tilde{\Sigma})$ (in this paper,  ${\cal F}(X)$ denotes the
family of all fuzzy subsets of $X$), and satisfy: For any $
\tilde{\sigma} \in \tilde{\Sigma}$,
\begin{equation}
\tilde{\Sigma}_{uc}(\tilde{\sigma})+\tilde{\Sigma}_{c}(\tilde{\sigma} )=1.
\end{equation}
Notably, the condition described by Eq. (5) is similar to that for
defining fuzzy observability by Lin and Ying [19]. For instance,
in the face of deciding to treat a patient having cancer via
either operation or drug therapy, either method (viewed as an
fuzzy event) has certain degree to be controlled. A sublanguage of
${\cal L}_{ \tilde{G}}$ is represented as $\tilde{K}\in {\cal
F}(\tilde{\Sigma}^{*})$ satisfying $\tilde{K}\subseteq {\cal L}_{
\tilde{G}}$. In this paper, $\tilde{A}\subseteq \tilde{B}$ stands
for $\tilde{A} (\tilde{\sigma}) \leq \tilde{B}(\tilde{\sigma})$
for any element $\tilde{\sigma}$ of domain. In the procedure of
curing a severe patient, a single physician-in-charge will be
confronted with some great challenges when making some crucial
decisions, in which some mistakes also possibly happen by
accident. In this regard, a supervisor (it may consist of a group
of specialists) likewise plays an important role. A  supervisor
$\tilde{S}$ of fuzzy DES $\tilde{G}$ is defined as a function:
 \[
\tilde{S}: \tilde{\Sigma}^{*}\rightarrow {\cal F}(\tilde{\Sigma}),
\]
where for each $\tilde{s}\in\tilde{\Sigma}^{*}$ and each
$\tilde{\sigma}\in \tilde{\Sigma}$,
$\tilde{S}(\tilde{s})(\tilde{\sigma})$ represents the possibility
of fuzzy event $\tilde{\sigma}$ being enabled after the occurrence
of fuzzy event string $\tilde{s}$, and $\min\{
\tilde{S}(\tilde{s})(\tilde{\sigma}), {\cal L}_{\tilde{G}}
(\tilde{s}\tilde{\sigma})\}$ is interpreted to be the degree to
which string $\tilde{s}\tilde{\sigma}$ is physically possible and
fuzzy event $\tilde{\sigma}$ is enabled after the occurrence of
fuzzy event string $\tilde{s}$. Similar to the admissibility
condition Eq. (4) indicated above for crisp supervisors,
$\tilde{S}$ is usually required to satisfy that for any
$\tilde{s}\in\tilde{\Sigma}^{*}$ and  $\tilde{\sigma}\in
\tilde{\Sigma}$,
\begin{equation}
\min\left\{\tilde{\Sigma}_{uc}(\tilde{\sigma}),{\cal L}_{\tilde{G}}(\tilde{s}\tilde{\sigma})\right\}\leq
\tilde{S}(\tilde{s})(\tilde{\sigma}).
\end{equation}
We may call this condition characterized by Eq. (6) as the {\it
fuzzy admissibility condition} for supervisor $\tilde{S}$ of fuzzy
DES $\tilde{G}$. To a great extent, this condition conforms to
real-life control problems. For example, in a therapeutic regime
for a patient having cancer, after a sequence of supervised
treatments, say $\tilde{s}$, a physician and a supervisor may face
the choice between operation $\tilde{\sigma}_{1}$ and drug therapy
$\tilde{\sigma}_{2}$. In a way, the possibilities of
$\tilde{\sigma}_{1}$ and  $\tilde{\sigma}_{2}$ being controlled
are small if there does not have any another therapy to be chosen
at present. Likely, $\tilde{\sigma}_{1}$ or $\tilde{\sigma}_{2}$
will be adopted following $\tilde{s}$, which means that for the
moment the degrees to which supervisor $\tilde{s}$ (group of
specialists) can control operation $\tilde{\sigma}_{1}$ and drug
therapy $\tilde{\sigma}_{2}$ are small, that is to say,
$\tilde{S}(\tilde{s})(\tilde{\sigma}_{i})$ $(i=1,2)$ that
represent the possibilities of  $\tilde{\sigma}_{1}$ and
$\tilde{\sigma}_{2}$ being controlled for the present are not
smaller than the possibility of $\tilde{s}\tilde{\sigma}_{i}$
being implemented and $\tilde{\sigma}_{i}$ uncontrollable.

In particular, if it is only required that Eq. (6) holds true for
any string  $ \tilde{s} $ with $ |\tilde{s}|\leq n$ (in this
paper, $|\tilde{s}|$ denotes the length of string $\tilde{s}$),
then we call it {\it fuzzy $n$-admissibility condition}.
Intuitively, the fuzzy admissibility condition Eq. (6) for
supervisor $\tilde{S}$ of fuzzy DES $\tilde{G}$ means that the
degree to which any fuzzy event $\tilde{\sigma}\in \tilde{\Sigma}$
following any string of fuzzy events
$\tilde{s}\in\tilde{\Sigma}^{*}$ is possible together with the
fuzzy event $\tilde{\sigma}$ being uncontrollable is not bigger
than the possibility for $\tilde{\sigma}$ being enabled after
$\tilde{s}$ occurring; as such, the fuzzy $n$-admissibility
condition for supervisor $\tilde{S}$ can be interpreted via
restricting the length of string of fuzzy events
$\tilde{s}\in\tilde{\Sigma}^{*}$ with $|\tilde{s}|\leq n$.

The fuzzy controlled system by $\tilde{S}$, denoted by $
\tilde{S}/\tilde{G} $, is also a fuzzy DES and the languages $
{\cal L}_{\tilde{S}/\tilde{G}} $ and ${\cal
L}_{\tilde{S}/\tilde{G},m}$ generated and marked by
$\tilde{S}/\tilde{G}$ respectively are defined as follows: For any
$\tilde{s}\in\tilde{\Sigma}^{*}$ and each $\tilde{\sigma}\in
\tilde{\Sigma}$,

$ {\cal L}_{\tilde{S}/\tilde{G}} (\epsilon)=1,\hskip
4mm {\cal
L}_{\tilde{S}/\tilde{G}}(\tilde{s}\tilde{\sigma})=\min\left\{{\cal
L}_{\tilde{S}/\tilde{G}}(\tilde{s}), {\cal
L}_{\tilde{G}}(\tilde{s}\tilde{\sigma}),\tilde{S}(\tilde{s})(\tilde{\sigma})\right\};
$

 $
{\cal L}_{\tilde{S}/\tilde{G},m} ={\cal L}_{\tilde{S}/\tilde{G}} \tilde{\cap} {\cal L}_{\tilde{G},m},
$\\
where symbol $\tilde{\cap}$ is Zadeh fuzzy AND operator, i.e.,
$(\tilde{A}\tilde{\cap}\tilde{B})(x)=\min\{\tilde{A}(x),\tilde{B}(x)\}$.
We give a notation concerning prefix-closed
property in the sense of fuzzy DESs. For any
$\tilde{s}\in\tilde{\Sigma}^{*}$,
\begin{equation}
pr(\tilde{s})=\{\tilde{t}\in\tilde{\Sigma}^{*}:\exists \tilde{r}\in \tilde{\Sigma}^{*}, \tilde{t}\tilde{r}=\tilde{s}\}.
\end{equation}
For any fuzzy language ${\cal L}$ over $\tilde{\Sigma}^{*}$, its
prefix-closure $pr({\cal L}):\tilde{\Sigma}^{*}\rightarrow [0,1]$
is defined as:
\begin{equation}
pr({\cal L})(\tilde{s})=\sup_{\tilde{s}\in pr(\tilde{t})}{\cal L}(\tilde{t}).
\end{equation}
So $pr({\cal L})(\tilde{s})$ denotes the possibility of string
$\tilde{s}$ belonging to the prefix-closure of ${\cal L}$. By
means of the formulation of the above concepts, now we can present
the controllability theorem concerning fuzzy DESs.

{\it Theorem 1.} Let a fuzzy DES be modeled by fuzzy finite
automaton (max-product or max-min automaton)
$\tilde{G}=(\tilde{Q},\tilde{\Sigma},\tilde{\delta},\tilde{q}_{0})$.
Suppose fuzzy uncontrollable subset $\tilde{\Sigma}_{uc}\in {\cal
F}(\tilde{\Sigma})$, and fuzzy legal subset $\tilde{K}\in {\cal
F}( \tilde{\Sigma}^{*} )$ that satisfies: $\tilde{K}\subseteq
{\cal L}_{\tilde{G}}$, and $\tilde{K}(\epsilon)=1$. Then there
exists supervisor $\tilde{S}: \tilde{\Sigma}^{*}\rightarrow {\cal
F}(\tilde{\Sigma})$,  such that $\tilde{S}$ satisfies the fuzzy
admissibility condition Eq. (6) and ${\cal
L}_{\tilde{S}/\tilde{G}}=pr(\tilde{K})$ if and only if  for any
$\tilde{s}\in \tilde{\Sigma}^{*}$ and any $\tilde{\sigma}\in
\tilde{\Sigma}$,
\begin{equation}
\min\left\{ pr(\tilde{K})(\tilde{s}),\tilde{\Sigma}_{uc}(\tilde{\sigma}),{\cal L}_{\tilde{G}}( \tilde{s}\tilde{\sigma} )\right\}
\leq  pr(\tilde{K})(\tilde{s}\tilde{\sigma}),
\end{equation}
where Eq. (9) is called {\it fuzzy controllability condition of
$\tilde{K}$ with respect to $\tilde{G}$ and $\tilde{\Sigma}_{uc}$}.

Intuitively, Eq. (9) means that the degree to which string
$\tilde{s}$ belongs to the prefix-closure of $\tilde{K}$ and fuzzy
event $\tilde{\sigma}$ following the string $\tilde{s}$ is
physically possible together with $ \tilde{\sigma} $ being
uncontrollable is not bigger than the possibility of the string
$\tilde{s}\tilde{\sigma} $ pertaining to  the prefix-closure of
$\tilde{K}$.

{\it Proof of Theorem 1:} First we note that  $\tilde{K} \subseteq {\cal
L}_{\tilde{G}}$ implies $pr(\tilde{K}) \subseteq {\cal
L}_{\tilde{G}}$. Indeed, for any $\tilde{s}\in \tilde{\Sigma}^{*}$, with Eq. (3) we have
\[
pr(\tilde{K})(\tilde{s})= \sup_{\tilde{t}\in\tilde{\Sigma}^{*}}\tilde{K} (\tilde{s}\tilde{t}) \leq
\sup_{\tilde{t}\in\tilde{\Sigma}^{*}} {\cal L}_{\tilde{G}} (\tilde{s}\tilde{t})={\cal L}_{\tilde{G}}(\tilde{s}),
\]
which verifies this result. We begin to show that if the fuzzy
controllability condition Eq. (9) holds, then there exists fuzzy
supervisor $\tilde{S}$ satisfying the required conditions. We
define $\tilde{S}: \tilde{\Sigma}^{*}\rightarrow {\cal
F}(\tilde{\Sigma})$ as: For any $\tilde{s}\in\tilde{\Sigma}^{*}$
and any $\tilde{\sigma}\in\tilde{\Sigma}$,
\begin{equation}
\tilde{S}(\tilde{s})(\tilde{\sigma})=\left\{\begin{array}{ll}
\min\left\{ \tilde{\Sigma}_{uc}(\tilde{\sigma}),{\cal
L}_{\tilde{G}}(\tilde{s}\tilde{\sigma})\right\}, & {\rm if}\hskip
2mm \tilde{\Sigma}_{uc}(\tilde{\sigma})\geq
pr(\tilde{K})(\tilde{s}\tilde{\sigma}),\\
pr(\tilde{K})(\tilde{s}\tilde{\sigma}),& {\rm otherwise}.
\end{array}
\right.
\end{equation}
Clearly, $\tilde{S}$ satisfies the fuzzy admissibility condition.
Next our purpose is to show that for any
$\tilde{s}\in\tilde{\Sigma}^{*}$,
\begin{equation}
{\cal L}_{\tilde{S}/\tilde{G}} (\tilde{s})= pr(\tilde{K}) (\tilde{s}).
\end{equation}
Proceed by induction for the length of $\tilde{s}$. If
$|\tilde{s}|=0$, i.e., $\tilde{s}=\epsilon$, then $ {\cal
L}_{\tilde{S}/\tilde{G}} (\epsilon)=1= pr(\tilde{K})(\epsilon)$.
Suppose that Eq. (11) holds true for any
$\tilde{s}\in\tilde{\Sigma}^{*}$ with $|\tilde{s}|\leq k-1$. Then
our aim is to prove that Eq. (11) holds for any
$\tilde{t}\in\tilde{\Sigma}^{*}$ with $|\tilde{t}|=k$. Let $
\tilde{t}= \tilde{s} \tilde{\sigma}$ where $|\tilde{s}|=k-1$. Then
with the assumption of induction, and the definition of ${\cal
L}_{\tilde{S}/\tilde{G}}$, we have
\begin{eqnarray*}
{\cal L}_{\tilde{S}/\tilde{G}} (\tilde{s}\tilde{\sigma} )&=&\min\left\{{\cal L}_{\tilde{S}/\tilde{G}} (\tilde{s}),
{\cal L}_{\tilde{G}}( \tilde{s}\tilde{\sigma} ), \tilde{S}(\tilde{s})(\tilde{\sigma})\right\}\\
&=&\min\left\{pr(\tilde{K})(\tilde{s}),{\cal L}_{\tilde{G}}(\tilde{s}\tilde{\sigma} ), \tilde{S}(\tilde{s})(\tilde{\sigma})\right\}.
\end{eqnarray*}
By means of the definition $\tilde{S}(\tilde{s})(\tilde{\sigma})$, if
$\tilde{\Sigma}_{uc}(\tilde{\sigma})\geq pr(\tilde{K})(\tilde{s}\tilde{\sigma} )$, then
\begin{eqnarray*}
{\cal L}_{\tilde{S}/\tilde{G}} (\tilde{s}\tilde{\sigma} ) &=&
\min\left\{ pr(\tilde{K})(\tilde{s}),{\cal L}_{\tilde{G}}(\tilde{s}\tilde{\sigma} ), \tilde{\Sigma}_{uc} ( \tilde{\sigma} )\right\}\\
&\leq& pr(\tilde{K})(\tilde{s}\tilde{\sigma} );
\end{eqnarray*}
if $\tilde{\Sigma}_{uc}(\tilde{\sigma})< pr(\tilde{K})(\tilde{s}\tilde{\sigma} )$, then
\begin{eqnarray*}
{\cal L}_{\tilde{S}/\tilde{G}} (\tilde{s}\tilde{\sigma} ) &=&
\min\left\{ pr(\tilde{K})(\tilde{s}),{\cal L}_{\tilde{G}}(\tilde{s}\tilde{\sigma} ), pr(\tilde{K})(\tilde{s}\tilde{\sigma}) \right\}\\
&\leq& pr(\tilde{K})(\tilde{s}\tilde{\sigma}).
\end{eqnarray*}
Therefore, we have shown ${\cal L}_{\tilde{S}/\tilde{G}}
(\tilde{s}\tilde{\sigma} ) \leq
pr(\tilde{K})(\tilde{s}\tilde{\sigma})$. On the other hand, due to
$pr(\tilde{K})(\tilde{s}\tilde{\sigma})\leq
pr(\tilde{K})(\tilde{s})$ and $ pr(\tilde{K})(
\tilde{s}\tilde{\sigma}) \leq {\cal
L}_{\tilde{G}}(\tilde{s}\tilde{\sigma})$, we have
\begin{eqnarray}
pr(\tilde{K})( \tilde{s}\tilde{\sigma})&\leq &\nonumber \min\left\{pr(\tilde{K})(\tilde{s}), {\cal L}_{\tilde{G}}(\tilde{s}\tilde{\sigma})\right\}\\
&=&\min\left\{ {\cal L}_{\tilde{S}/\tilde{G}} (\tilde{s}),{\cal L}_{\tilde{G}}(\tilde{s}\tilde{\sigma})\right\}.
\end{eqnarray}
Furthermore, if  $ \tilde{\Sigma}_{uc}(\tilde{\sigma})\geq
pr(\tilde{K})(\tilde{s}\tilde{\sigma} )$, then by combining the
definition of $\tilde{S}(\tilde{s})$ with Eq. (12), we have
\begin{eqnarray*}
pr(\tilde{K})( \tilde{s}\tilde{\sigma}) &\leq &\min\left\{{\cal L}_{\tilde{S}/\tilde{G}} (\tilde{s}),
 {\cal L}_{\tilde{G}}(\tilde{s}\tilde{\sigma}),   \tilde{\Sigma}_{uc}(\tilde{\sigma})
\right\}\\
&=&\min\left\{ {\cal L}_{\tilde{S}/\tilde{G}} (\tilde{s}),{\cal L}_{\tilde{G}} (\tilde{s}\tilde{\sigma}), \tilde{S}(\tilde{s})(\tilde{\sigma})\right\}\\
&=&{\cal L}_{\tilde{S}/\tilde{G}} (\tilde{s}\tilde{\sigma});
\end{eqnarray*}
if  $ \tilde{\Sigma}_{uc}(\tilde{\sigma})< pr(\tilde{K})
(\tilde{s}\tilde{\sigma} )$, then with the definition of
$\tilde{S}(\tilde{s})$ we have
$\tilde{S}(\tilde{s})(\tilde{\sigma})=pr(\tilde{K})(\tilde{s}\tilde{\sigma}
)$, and by Eq. (12) we obtain that
\begin{eqnarray*}
pr(\tilde{K})( \tilde{s}\tilde{\sigma})&\leq & \min\left\{ {\cal
L}_{\tilde{S}/\tilde{G}} (\tilde{s}), {\cal L}_{\tilde{G}}
(\tilde{s}\tilde{\sigma}),
\tilde{S}(\tilde{s})(\tilde{\sigma}) \right\}\\
&=& {\cal L}_{\tilde{S}/\tilde{G}} (\tilde{s}\tilde{\sigma}).
\end{eqnarray*}
Therefore we have verified that $ pr(\tilde{K})(
\tilde{s}\tilde{\sigma})={\cal L}_{\tilde{S}/\tilde{G}}
(\tilde{s}\tilde{\sigma})$ holds for any
$\tilde{s}\in\tilde{\Sigma}^{*}$ and $\tilde{\sigma}\in
\tilde{\Sigma}$ with $|\tilde{s}|=k-1$, and the proof of
sufficiency is completed.

The remainder is to consider the  proof of necessity. If ${\cal
L}_{\tilde{S}/\tilde{G}}=pr(\tilde{K})$ holds, then for any
$\tilde{s}\in\tilde{\Sigma}^{*}$ and $\tilde{\sigma}\in
\tilde{\Sigma}$, with the fuzzy admissibility condition Eq. (6) of
$\tilde {S}$ we have
\begin{eqnarray*}
&&\min\left\{pr(\tilde{K})( \tilde{s}),\tilde{\Sigma}_{uc}(\tilde{\sigma}), {\cal L}_{\tilde{G}} (\tilde{s}\tilde{\sigma})\right\}\\
&\leq&\min\left\{ {\cal L}_{\tilde{S}/\tilde{G}} (\tilde{s}), \tilde{S}(\tilde{s})(\tilde{\sigma}),
{\cal L}_{\tilde{G}} (\tilde{s}\tilde{\sigma})\right\}\\
&=&{\cal L}_{\tilde{S}/\tilde{G}} (\tilde{s}\tilde{\sigma})\\
&=&pr(\tilde{K})( \tilde{s}\tilde{\sigma}).
\end{eqnarray*}
This completes the proof of necessity and the theorem has therefore been proved. $\Box$

{\it Remark 4.} Theorem 1 together with its proof presents a
method for designing a supervisor of a fuzzy DES such that the
deriving fuzzy system obeys the pre-specified ``legal" behavior
for the fuzzy controlled system, which may apply to fuzzy DESs
such as modeling a patient's health condition and traffic systems.
In particular, from the above proof we see that in Theorem 1 the
fuzzy finite automaton can be of max-product or max-min. We will
see that when fuzzy DES is modeled by max-product automaton, it is
difficult to check the fuzzy controllability condition Eq. (9)
with a general criterion, whereas if max-min automata are used to
model fuzzy DESs, then we can present a general computing process
in detail for testing the fuzzy controllability condition Eq. (9).

As an aspect of application, we can utilize Theorem 1 to cope with
some realistic control problems with finite length specifications.
 In this regard, from Theorem 1 it follows readily
the following Corollary 1. Before giving this corollary, we
introduce two notations: For any two fuzzy languages $\tilde{A}$
and $\tilde{B}$ over set $\tilde{\Sigma}$ of fuzzy events,
$\tilde{A}\subseteq_{n}\tilde{B}$ and $\tilde{A}=_{n}\tilde{B}$
mean respectively that for any string $ \tilde{s}
\in\tilde{\Sigma}^{*}$ with $|\tilde{s}|\leq n$,
$\tilde{A}(\tilde{s})\leq\tilde{B}(\tilde{s})$ and
$\tilde{A}(\tilde{s})=\tilde{B}(\tilde{s})$.

{\it Corollary 1.} Let a fuzzy DES be modeled by fuzzy finite
automaton (max-product or max-min)
$\tilde{G}=(\tilde{Q},\tilde{\Sigma},\tilde{\delta},\tilde{q}_{0})$.
Let $n$ be any given positive integer. Suppose fuzzy
uncontrollable subset $\tilde{\Sigma}_{uc}\in {\cal
F}(\tilde{\Sigma})$, and fuzzy legal subset $\tilde{K}\in {\cal
F}( \tilde{\Sigma}^{*} )$ that satisfies: $\tilde{K}\subseteq_{n}
{\cal L}_{\tilde{G}}$, and $\tilde{K}(\epsilon)=1$. Then there
exists supervisor $\tilde{S}: \tilde{\Sigma}^{*}\rightarrow {\cal
F}(\tilde{\Sigma})$,  such that $\tilde{S}$ satisfies the fuzzy
$n$-admissibility condition Eq. (6) and ${\cal
L}_{\tilde{S}/\tilde{G}}=_{n} pr(\tilde{K})$ if and only if  for
any $\tilde{s}\in \tilde{\Sigma}^{*}$ with $|\tilde{s}|\leq n$ and
any $\tilde{\sigma}\in \tilde{\Sigma}$,
\begin{equation}
\min\left\{ pr(\tilde{K})(\tilde{s}),\tilde{\Sigma}_{uc}(\tilde{\sigma}),{\cal L}_{\tilde{G}}( \tilde{s}\tilde{\sigma} )\right\}
\leq  pr(\tilde{K})(\tilde{s}\tilde{\sigma}),
\end{equation}
where Eq. (13) is called {\it fuzzy $n$-controllability condition of
$\tilde{K}$ with respect to $\tilde{G}$ and $\tilde{\Sigma}_{uc}$}.

{\it Proof:} Exactly similar to Theorem 1 by restricting the length of $\tilde{s}$ with $|\tilde{s}|\leq n$. $\Box$

{\it Remark 5.}  If for every fuzzy event $\tilde{\sigma}\in
\tilde{\Sigma}$ and every string $\tilde{s}\in
\tilde{\Sigma}^{*}$, we view the total computation for Eq. (13) as
a step,   then with Eq. (13) it is possible to check whether or
not {\it fuzzy $n$-controllability condition} holds with most
number of computing steps $(1+ |\tilde{\Sigma}|
+2^{|\tilde{\Sigma}| }+\cdots +n^{|\tilde{\Sigma}|
})|\tilde{\Sigma}| $,  where $ |\tilde{\Sigma}|$  is the number of
fuzzy events. Therefore the worst-case computational complexity
for this test is exponential, and it should be worth considering
to reduce the computation procedure.

\subsection*{{\it B. Realization of Supervisors for Fuzzy DESs and Some Examples}}

From Theorem 1 it is seen that the existence of supervisor
$\tilde{S}$ is closely related to the fuzzy controllability
condition of $\tilde{K}$ defined by Eq. (9), so it is very
important to consider how to test such a condition. Nevertheless,
in fuzzy DESs modeled by max-product automata, the set of fuzzy
states $\{\tilde{q}_{0}\circ \tilde{s}:\tilde{s}\in
\tilde{\Sigma}^{*}\}$ is likely infinite. Along with the further
analysis in the following (Case 1), it is quite complicated to
present a uniform method to check the fuzzy controllability
condition for this case. But when the fuzzy DESs modeled by
max-min automata, we can give a general computing flow for testing
the condition (Case 2). The two cases are dealt with as follows,
and we focus on the second case.

{\it Case 1.} {\it Fuzzy DESs modeled by max-product automata.}

In fuzzy DES modeled by max-product automaton
$\tilde{G}=(\tilde{Q},\tilde{\Sigma},\tilde{\delta},\tilde{q}_{0})$,
in which the dimensionality of $\tilde{Q}$ is $n$ (i.e., $Q$ has
$n$ crisp states), for a given fuzzy subset $\tilde{K}\subseteq
{\cal L}_{\tilde{G}}$  of control specifications, by means of
Theorem 1 we should decide the existence of supervisor
$\tilde{S}$. As in crisp DESs [1], we may assume $pr(\tilde{K})$
to be a fuzzy language generated by a max-product automaton
$\tilde{H}=(\tilde{R},\tilde{\Sigma},\tilde{\gamma},\tilde{p_{0}})$
with dimensionality $m$ of $\tilde{R}$. Then we have that for any
$\tilde{s}=
\tilde{\sigma}_{1}\tilde{\sigma}_{2}\cdots\tilde{\sigma}_{k} \in
\tilde{\Sigma}^{*}$ where $\tilde{\sigma}_{i}\in\tilde{\Sigma}$,
$i=1,2,\cdots,k$,
\begin{equation}
pr(\tilde{K})(\tilde{s}) ={\cal L}_{\tilde{H}}(\tilde{s})=\max_{i=1}^{m}\tilde{p_{0}}\circ
\tilde{\sigma}_{1}\circ\tilde{\sigma}_{2}\circ\cdots\circ\tilde{\sigma}_{k} \circ \bar{s}_{i}^{T},
\end{equation}
\begin{equation}
pr(\tilde{K})(\tilde{s}\tilde{\sigma}) = {\cal L}_{\tilde{H}}(\tilde{s}\tilde{\sigma}) = \max_{i=1}^{m}\tilde{p_{0}}\circ
\tilde{\sigma}_{1}\circ\tilde{\sigma}_{2}\circ\cdots\circ\tilde{\sigma}_{k} \circ\tilde{\sigma}\circ \bar{s}_{i}^{T},
\end{equation}
\begin{equation}
{\cal L}_{\tilde{G}}(\tilde{s}\tilde{\sigma}) =\max_{i=1}^{m}\tilde{q_{0}}\circ
\tilde{\sigma}_{1}\circ\tilde{\sigma}_{2}\circ\cdots\circ\tilde{\sigma}_{k} \circ\tilde{\sigma}\circ \bar{s}_{i}^{T},
\end{equation}
where $ \bar{s}_{i}^{T} $ $(i=1,2,\ldots,m)$ and $
\bar{s}_{j}^{T}$ $(j=1,2,\ldots,m)$, as indicated above, are
respectively the crisp states of $\tilde{H}$ and $\tilde{G}$. If
the sets of fuzzy states
$\{\tilde{p_{0}}\circ\tilde{s}:\tilde{s}\in \tilde{\Sigma}^{*}\}$
and
$\{\tilde{q_{0}}\circ\tilde{s}:\tilde{s}\in\tilde{\Sigma}^{*}\}$
were finite, then by virtue of Eqs. (14-16) we can check the fuzzy
controllability condition described above by Eq. (9) within finite
steps. But unfortunately it is perhaps not so. From Eqs. (14) and
(15) it follows that $pr(\tilde{K})(\tilde{s})$ and
$pr(\tilde{K})(\tilde{s}\tilde{\sigma})$  may be arbitrarily small
in case of the length of $\tilde{s}$ big enough, since those
elements in fuzzy events (matrices) are usually smaller than 1. On
the other hand, $pr(\tilde{K})(\tilde{s}\tilde{\sigma})\leq {\cal
L}_{\tilde{G}}(\tilde{s}\tilde{\sigma}) $ is prerequisite, and
$pr(\tilde{K})(\tilde{s}\tilde{\sigma})\leq
pr(\tilde{K})(\tilde{s})$ always holds, that is, we always have
\begin{equation}
\min\{ pr(\tilde{K})(\tilde{s}\tilde{\sigma}), \tilde{\Sigma}_{uc}(\tilde{\sigma})\}\leq
\min\left\{ pr(\tilde{K})(\tilde{s}),\tilde{\Sigma}_{uc}(\tilde{\sigma}),{\cal L}_{\tilde{G}}( \tilde{s}\tilde{\sigma} )\right\};
\end{equation}
however, $\tilde{\Sigma}_{uc}(\tilde{\sigma})$ is certain for each
fuzzy event $\tilde{\sigma}$, and, as just indicated,
$pr(\tilde{K})(\tilde{s})$ and
$pr(\tilde{K})(\tilde{s}\tilde{\sigma}) $  tend likely to zero
when $|\tilde{s}|$ is big enough, and it thus follows from Eq.
(17) that
\[
pr(\tilde{K})(\tilde{s}\tilde{\sigma})\leq
\min\left\{ pr(\tilde{K})(\tilde{s}),\tilde{\Sigma}_{uc}(\tilde{\sigma}),{\cal L}_{\tilde{G}}( \tilde{s}\tilde{\sigma} )\right\},
\]
which implies that to guarantee the fuzzy
controllability condition Eq. (9), it is required that as the length of string $\tilde{s}$ becomes big enough,
\begin{equation}
pr(\tilde{K})( \tilde{s}\tilde{\sigma})=
\min\left\{ pr(\tilde{K})(\tilde{s}),{\cal L}_{\tilde{G}}( \tilde{s}\tilde{\sigma} )\right\}.
\end{equation}
Clearly, Eq. (18) is a strict condition on $\tilde{K}$, and,
therefore, to ensure the fuzzy
controllability condition Eq. (9), some restrictions imposed upon $\tilde{K}$ are necessary. For example,
\[
\tilde{K}( \tilde{s}\tilde{\sigma} ) \geq \min \{
\tilde{\Sigma}_{uc}(\tilde{\sigma}), {\cal
L}_{\tilde{G}}(\tilde{s}\tilde{\sigma} )\}
\]
is a sufficient condition to result in Eq. (9); or when the
support set of $\tilde{K}$ is finite, we can check whether or not
Eq. (9) holds.

{\it Case 2.} {\it Fuzzy DESs modeled by max-min automata.}

Let fuzzy DES be modeled by max-min automaton
$\tilde{G}=(\tilde{Q},\tilde{\Sigma},\tilde{\delta},\tilde{q}_{0})$,
in which the dimensionality of $\tilde{Q}$ is $n$. We first give a
{\it computing tree} for deriving the set of all fuzzy states
reachable from the initial state $\tilde{q}_{0}$, and the sets of
strings respectively corresponding to each accessible fuzzy state
are also obtained. Assume that
$\tilde{\Sigma}=\{\tilde{\alpha}_{1},\tilde{\alpha}_{2},\ldots,\tilde{\alpha}_{n}\}$.
A basic idea is based on that (i)
$\tilde{q}_{0}\odot\tilde{s}=\tilde{q}_{0}\odot\tilde{s}\odot
(\tilde{s}_{1})^{n}$ for any $n\geq 0$ if
$\tilde{q}_{0}\odot\tilde{s}=\tilde{q}_{0}\odot\tilde{s}\odot
\tilde{s}_{1}$ for $ \tilde{s}_{1}\in \tilde{\Sigma}^{*}$, where
$(\tilde{s}_{1})^{n}$ denotes the $\odot$ product of $n$'s
$\tilde{s}_{1}$, and (ii) the set of fuzzy states
$\{\tilde{q}_{0}\odot\tilde{s}:\tilde{s}\in\tilde{\Sigma}^{*}\}$
is always finite since $\tilde{\Sigma}$ is finite. For the sake of
simplicity, we present the computing tree for
$\tilde{\Sigma}=\{\tilde{\alpha}_{1},\tilde{\alpha}_{2}\}$ of two
fuzzy events via Fig. 3, and the case of more than two fuzzy
events is analogous.

\setlength{\unitlength}{0.05in}

\begin{picture}(60,60)

\put(50,61){\makebox(10,5)[c]{{\tiny Begin}}}

\put(55,62){\vector(0,-1){3}}

\put(50,55){\makebox(10,5)[c]{{\tiny $\tilde{q}_{0}$}}}

\put(55,55){\line(0,-1){5}} \put(50,50){\line(1,0){10}}
\put(50,50){\vector(0,-1){5}}\put(60,50){\vector(0,-1){5}}

\put(45,40){\makebox(10,5)[c]{{\tiny
$\tilde{q}_0\odot\tilde{\alpha}_1$}}}\put(55,40){\makebox(10,5)[c]{{\tiny
$\tilde{q}_0\odot\tilde{\alpha}_2$}}}
\put(50,40){\vector(0,-1){10}}\put(25,35){\line(1,0){25}}
\put(25,35){\vector(0,-1){5}}
\put(60,40){\vector(0,-1){10}}\put(60,35){\line(1,0){25}}
\put(85,35){\vector(0,-1){5}} \put(15,25){\makebox(10,5)[c]{{\tiny
$\tilde{q}_0\odot\tilde{\alpha}_1\odot\tilde{\alpha}_1$}}}\put(43,25){\makebox(10,5)[c]{{\tiny
$\tilde{q}_0\odot\tilde{\alpha}_1\odot\tilde{\alpha}_2$}}}
\put(57,25){\makebox(10,5)[c]{{\tiny
$\tilde{q}_0\odot\tilde{\alpha}_2\odot\tilde{\alpha}_1$}}}\put(85,25){\makebox(10,5)[c]{{\tiny
$\tilde{q}_0\odot\tilde{\alpha}_2\odot\tilde{\alpha}_2$}}}

\put(15,25){\vector(0,-1){10}}\put(0,20){\line(1,0){15}}
\put(0,20){\vector(0,-1){5}}
\put(50,25){\vector(0,-1){10}}\put(25,20){\line(1,0){25}}
\put(25,20){\vector(0,-1){5}}

\put(60,25){\vector(0,-1){10}}\put(60,20){\line(1,0){25}}
\put(85,20){\vector(0,-1){5}}
\put(95,25){\vector(0,-1){10}}\put(95,20){\line(1,0){15}}
\put(110,20){\vector(0,-1){5}}
\put(-5,10){\makebox(10,5)[c]{{\tiny
$\tilde{q}_0\odot\tilde{\alpha}_1^3$}}}
\put(7,10){\makebox(10,5)[c]{{\tiny
$\tilde{q}_0\odot\tilde{\alpha}_1^2\odot\tilde{\alpha}_2$}}}
\put(32,10){\makebox(-5,5)[c]{{\tiny
$\tilde{q}_0\odot\tilde{\alpha}_1\odot\tilde{\alpha}_2\odot\tilde{\alpha}_1$}}}
\put(44,10){\makebox(6,5)[c]{{\tiny
$\tilde{q}_0\odot\tilde{\alpha}_1\odot\tilde{\alpha}_2^2$}}}
\put(56,10){\makebox(10,5)[c]{{\tiny
$\tilde{q}_0\odot\tilde{\alpha}_2\odot\tilde{\alpha}_1^2$}}}
\put(72,10){\makebox(20,5)[c]{{\tiny
$\tilde{q}_0\odot\tilde{\alpha}_2\odot\tilde{\alpha}_1\odot\tilde{\alpha}_2$}}}
\put(93,10){\makebox(12,5)[c]{{\tiny
$\tilde{q}_0\odot\tilde{\alpha}_2^2\odot\tilde{\alpha}_1$}}}
\put(108,10){\makebox(10,5)[c]{{\tiny
$\tilde{q}_0\odot\tilde{\alpha}_2^3$}}}

\put(-5,7){\makebox(10,5)[c]{{\vdots}}}
\put(7,7){\makebox(10,5)[c]{{\vdots}}}
\put(20,7){\makebox(10,5)[c]{{\vdots}}}
\put(44,7){\makebox(10,5)[c]{{\vdots}}}
\put(56,7){\makebox(10,5)[c]{{\vdots}}}
\put(77,7){\makebox(10,5)[c]{{\vdots}}}
\put(95,7){\makebox(10,5)[c]{{\vdots}}}
\put(107,7){\makebox(10,5)[c]{{\vdots}}}

\put(45,45){\makebox(5,5)[c]{{\tiny
$\tilde{\alpha}_1$}}}\put(50,40){\makebox(25,15)[c]{{\tiny
$\tilde{\alpha}_2$}}} \put(15,25){\makebox(25,15)[c]{{\tiny
$\tilde{\alpha}_1$}}}\put(35,25){\makebox(25,15)[c]{{\tiny
$\tilde{\alpha}_2$}}} \put(50,25){\makebox(25,15)[c]{{\tiny
$\tilde{\alpha}_1$}}}\put(70,25){\makebox(25,15)[c]{{\tiny
$\tilde{\alpha}_2$}}} \put(35,10){\makebox(25,15)[c]{{\tiny
$\tilde{\alpha}_2$}}}\put(15,10){\makebox(25,15)[c]{{\tiny
$\tilde{\alpha}_1$}}} \put(50,10){\makebox(25,15)[c]{{\tiny
$\tilde{\alpha}_1$}}}\put(60,10){\makebox(45,15)[c]{{\tiny
$\tilde{\alpha}_2$}}} \put(-10,10){\makebox(25,15)[c]{{\tiny
$\tilde{\alpha}_1$}}}\put(0,10){\makebox(25,15)[c]{{\tiny
$\tilde{\alpha}_2$}}} \put(85,10){\makebox(25,15)[c]{{\tiny
$\tilde{\alpha}_1$}}}\put(95,10){\makebox(25,15)[c]{{\tiny
$\tilde{\alpha}_2$}}}

\put(50,0){\makebox(20,5)[c]{{\footnotesize Fig. 3. A computing
tree for deciding the all different fuzzy states reachable from
$\tilde{q}_{0}$. }}}

\end{picture}

\noindent In this computing tree, the initial fuzzy state
$\tilde{q}_{0}$ is its root; each vertex, say
$\tilde{q}_{0}\odot\tilde{s}$, may produce $n$'s sons
$\tilde{q}_{0}\odot\tilde{s}\odot\tilde{s}_{1}$,
$\tilde{q}_{0}\odot\tilde{s}\odot\tilde{s}_{2}$, $\ldots$,
$\tilde{q}_{0}\odot\tilde{s}\odot\tilde{s}_{1}$. However, if
$\tilde{q}_{0}\odot\tilde{s}$ equals some its father, then
$\tilde{q}_{0}\odot\tilde{s}$ is a leaf, that is marked by a
underline. The computing ends with a leaf at the end of each
branch.

{\it Example 2.} In max-min automaton
$\tilde{G}=(\tilde{Q},\tilde{\Sigma},\tilde{\delta},\tilde{q}_{0})$,
where $\tilde{\Sigma}=\{\tilde{\alpha}_{1},\tilde{\alpha}_{2}\}$,
$\tilde{q}_{0}=[0.9,\hskip 2mm 0.1]$, and
$\tilde{\alpha}_{1}=\left[\begin{array}{cc}
0.4&0.8\\
0.2&0.2
\end{array}
\right]$, $\tilde{\alpha}_{2}=\left[\begin{array}{cc}
0.4&0.2\\
0.8&0.5
\end{array}
\right]$. Then we have the following computing tree, in which each
leaf is underlined.

\setlength{\unitlength}{0.05in}

\begin{picture}(50,83)

\put(40,95){\makebox(10,5)[c]{{\tiny Begin}}}

\put(45,96){\vector(0,-1){3}}

\put(45,89){\makebox(10,5)[c]{{\tiny $\tilde{q}_0$=[0.9 0.1]}}}

\put(45,90){\line(0,-1){5}} \put(40,85){\line(1,0){10}}
\put(40,85){\vector(0,-1){5}}\put(50,85){\vector(0,-1){5}}
\put(35,75){\makebox(10,5)[c]{{\tiny [0.4 0.8]
}}}\put(45,75){\makebox(10,5)[c]{{\tiny [0.4 0.2]}}}
\put(40,75){\vector(0,-1){10}}\put(15,70){\line(1,0){25}}
\put(15,70){\vector(0,-1){5}}
\put(50,75){\vector(0,-1){10}}\put(50,70){\line(1,0){25}}
\put(75,70){\vector(0,-1){5}} \put(10,60){\makebox(10,5)[c]{{\tiny
[0.4 0.4]} }}\put(35,60){\makebox(10,5)[c]{{\tiny [0.8 0.5]}}}
\put(46,60){\makebox(10,5)[c]{{\tiny [0.4 0.4]}
}}\put(70,60){\makebox(10,5)[c]{{\tiny $\underline{[0.4\hskip 1mm
0.2]}$}}}

\put(15,60){\vector(0,-1){10}}\put(0,55){\line(1,0){15}}
\put(0,55){\vector(0,-1){5}}
\put(40,60){\vector(0,-1){10}}\put(25,55){\line(1,0){15}}
\put(25,55){\vector(0,-1){5}}

\put(50,60){\vector(0,-1){10}}\put(50,55){\line(1,0){15}}
\put(65,55){\vector(0,-1){5}}

\put(-5,45){\makebox(10,5)[c]{{\tiny $\underline{[0.4\hskip 1mm
0.4]}$}}} \put(10,45){\makebox(10,5)[c]{{\tiny
$\underline{[0.4\hskip 1mm 0.4]}$}}}
\put(20,45){\makebox(10,5)[c]{{\tiny $\underline{[0.4\hskip 1mm
0.8]}$}}} \put(35,45){\makebox(10,5)[c]{{\tiny [0.5 0.5]}}}
\put(45,45){\makebox(10,5)[c]{{\tiny $\underline{[0.4\hskip 1mm
0.4]}$}}} \put(60,45){\makebox(10,5)[c]{{\tiny
$\underline{[0.4\hskip 1mm 0.4]}$}}}

\put(35,80){\makebox(5,5)[c]{{\tiny
$\tilde{\alpha}_1$}}}\put(50,80){\makebox(5,5)[c]{{\tiny
$\tilde{\alpha}_2$}}} \put(15,65){\makebox(5,5)[c]{{\tiny
$\tilde{\alpha}_1$}}}\put(35,65){\makebox(5,5)[c]{{\tiny
$\tilde{\alpha}_2$}}} \put(50,65){\makebox(5,5)[c]{{\tiny
$\tilde{\alpha}_1$}}}\put(70,65){\makebox(5,5)[c]{{\tiny
$\tilde{\alpha}_2$}}} \put(25,50){\makebox(5,5)[c]{{\tiny
$\tilde{\alpha}_1$}}}\put(35,50){\makebox(5,5)[c]{{\tiny
$\tilde{\alpha}_2$}}} \put(50,50){\makebox(5,5)[c]{{\tiny
$\tilde{\alpha}_1$}}}\put(60,50){\makebox(5,5)[c]{{\tiny
$\tilde{\alpha}_2$}}} \put(0,50){\makebox(5,5)[c]{{\tiny
$\tilde{\alpha}_1$}}}\put(10,50){\makebox(5,5)[c]{{\tiny
$\tilde{\alpha}_2$}}} \put(40,45){\vector(0,-1){10}}
\put(25,40){\line(1,0){15}}

\put(25,40){\vector(0,-1){5}}

\put(25,35){\makebox(5,5)[c]{{\tiny
$\tilde{\alpha}_1$}}}\put(35,35){\makebox(5,5)[c]{{\tiny
$\tilde{\alpha}_2$}}}

\put(40,45){\vector(0,-1){10}}\put(25,40){\line(1,0){15}}

\put(25,40){\vector(0,-1){5}}

\put(20,30){\makebox(10,5)[c]{{\tiny [0.4 0.4]}}}

\put(35,30){\makebox(10,5)[c]{{\tiny $\underline{[0.5\hskip 1mm
0.5]}$}}}

\put(25,30){\vector(0,-1){10}}\put(10,25){\line(1,0){15}}
\put(10,25){\vector(0,-1){5}} \put(5,15){\makebox(10,5)[c]{{\tiny
$\underline{ [0.4\hskip 1mm 0.4]}$}}}
\put(20,15){\makebox(10,5)[c]{{\tiny $\underline{ [0.5\hskip 1mm
0.5]}$}}} \put(10,20){\makebox(5,5)[c]{{\tiny
$\tilde{\alpha}_1$}}}\put(20,20){\makebox(5,5)[c]{{\tiny
$\tilde{\alpha}_2$}}}

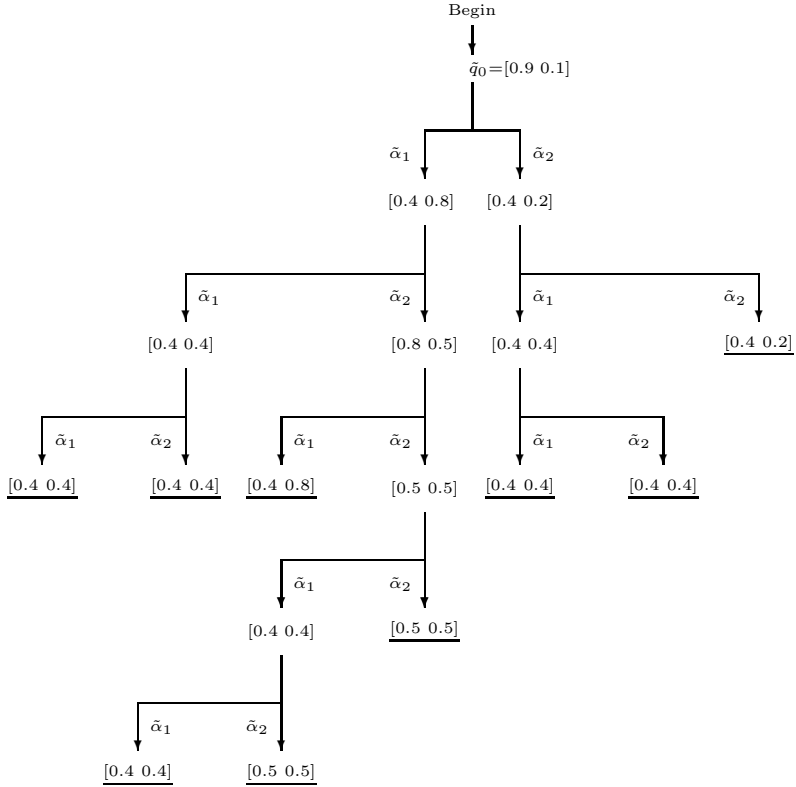
\put(30,10){\makebox(45,5)[c]{{\footnotesize Fig. 4.  A computing
tree for deciding the all different fuzzy states reachable from
$\tilde{q}_0$=[0.9 0.1].}}}

\end{picture}

\vskip -15mm

\noindent From this computing tree (Fig. 4) it follows that in
$\tilde{G}$ there are only six different fuzzy states reachable
from $\tilde{q}_{0}$=[0.9 0.1], which are listed in TABLE I as
follows.

\begin{picture}(50,5)

\put(20,1){\makebox(10,5)[c]{{\footnotesize TABLE  I }}}

\put(20,-1.5){\makebox(10,5)[c]{{\footnotesize  Six different
fuzzy states reachable from $\tilde{q}_{0}$.}}}

\end{picture}

\begin {tabular}{|c|c|c|c|}
\hline $\tilde{s}$&$\tilde{q}_0\odot \tilde{s}
$&$\tilde{s}$&$\tilde{q}_0\odot\tilde{s}$\\
\hline $\tilde{\epsilon}$&[0.9 \hskip 1mm 0.1]
&$\tilde{\alpha}_1\tilde{\alpha}_1$&[0.4 \hskip 1mm 0.4]\\
\hline $\tilde{\alpha}_1$&[0.4 \hskip 1mm
0.8]&$\tilde{\alpha}_1\tilde{\alpha}_2$&[0.8
\hskip 1mm 0.5]\\
\hline
 $\tilde{\alpha}_2$&[0.4 \hskip
1mm
0.2]&$\tilde{\alpha}_1\tilde{\alpha}_2\tilde{\alpha}_2$&[0.5 \hskip 1mm  0.5]\\
\hline

\end{tabular}

\noindent Also, the corresponding sets of strings to each fuzzy
states can be obtained respectively as follows:
$C(\tilde{q}_{0})=\{\epsilon\}$;
$C(\tilde{q}_{0}\odot\tilde{\alpha}_{1})=\{\tilde{\alpha}_{1}(\tilde{\alpha}_{2}\tilde{\alpha}_{1})^{n}:n\geq
0\}$;
$C(\tilde{q}_{0}\odot\tilde{\alpha}_{2})=\{\tilde{\alpha}_{2}^{n}:n\geq
1\}$;
\begin{eqnarray*}
C(\tilde{q}_{0}\odot\tilde{\alpha}_{1}^{2})&=&\{\tilde{\alpha}_{1}^{2}\tilde{\alpha}_{1}^{n}:n\geq 0\}\cup \{\tilde{\alpha}_{1}^{2}\tilde{\alpha}_{2}^{n}:n\geq 0\}\\
&&\cup\{\tilde{\alpha}_{1}\tilde{\alpha}_{2}\tilde{\alpha}_{2}\tilde{\alpha}_{1}^{n}:n\geq 1\}\cup \{\tilde{\alpha}_{1}\tilde{\alpha}_{2}\tilde{\alpha}_{2}\tilde{\alpha}_{1}\tilde{\alpha}_{2}^{n}:n\geq 0\}\\
&&\cup\{\tilde{\alpha}_{2}\tilde{\alpha}_{1}^{n}:n\geq
1\}\cup\{\tilde{\alpha}_{2}\tilde{\alpha}_{1}\tilde{\alpha}_{2}^{n}:n\geq
0\};
\end{eqnarray*}
$C(\tilde{q}_{0}\odot\tilde{\alpha}_{1}\odot\tilde{\alpha}_{2})=\{(\tilde{\alpha}_{1}\tilde{\alpha}_{2})^{n}:n\geq
1\}$; and
$C(\tilde{q}_{0}\odot\tilde{\alpha}_{1}\odot\tilde{\alpha}_{2}^{2})=\{\tilde{\alpha}_{1}\tilde{\alpha}_{2}\tilde{\alpha}_{2}^{n}:n\geq
1\}$. Here, $C(\tilde{q})$ denotes the set
$\{\tilde{s}\in\tilde{\Sigma}^{*}:
\tilde{\delta}(\tilde{q}_{0},\tilde{s})=
\tilde{q}_{0}\odot\tilde{s}=\tilde{q}\}$.

For two max-min automata
$\tilde{G}_{i}=(\tilde{Q}_{i},\tilde{\Sigma}_{i},\tilde{\delta}_{i},\tilde{q}_{i0})$
$(i=1,2)$  with the same set of fuzzy events
$\tilde{\Sigma}_{1}=\tilde{\Sigma}_{2}=\tilde{\Sigma}=\{\tilde{\sigma}_{1},\tilde{\sigma}_{2},\ldots,\tilde{\sigma}_{n}\}$,
our purpose is to search for the all different pairs of fuzzy
states reachable from the initial fuzzy state pair
$(\tilde{q}_{10},\tilde{q}_{20})$, that is,
$\{(\tilde{q}_{10}\odot\tilde{s},\tilde{q}_{20}\odot\tilde{s}):\tilde{s}\in\tilde{\Sigma}^{*}\}$.
The method is similar to the case of single max-min automaton
presented above, that is carried out by a computing tree. In the
computing tree, the root is labeled with pair
$(\tilde{q}_{10},\tilde{q}_{20})$, and each vertex, say
$(\tilde{q}_{10}\odot\tilde{s},\tilde{q}_{20}\odot\tilde{s})$ for
$\tilde{s}\in\tilde{\Sigma}^{*}$, may produce $n$'s sons, i.e.,
$(\tilde{q}_{10}\odot\tilde{s}\odot\tilde{\sigma}_{i},\tilde{q}_{20}\odot\tilde{s}\odot\tilde{\sigma}_{i})$,
$i=1,2,\ldots,n$. But if a pair
$(\tilde{q}_{10}\odot\tilde{s}\odot\tilde{\sigma}_{i},\tilde{q}_{20}\odot\tilde{s}\odot\tilde{\sigma}_{i})$
is the same as one of its father, then this pair will be treated
as a leaf, that is marked with a underline. Such a computing tree
is depicted by Fig. 5 as follows. Since the set of all pairs of
fuzzy states is finite due to the finiteness of $\tilde{\Sigma}$,
the computing tree ends with a leaf at the end of each branch.

\setlength{\unitlength}{0.05in}

\begin{picture}(60,60)

\put(50,56){\makebox(10,5)[c]{{\tiny Begin}}}
 \put(55,57){\vector(0,-1){3}}

 \put(50,50){\makebox(10,5)[c]{{\tiny
($\tilde{q}_{10}$\hskip 4mm $\tilde{q}_{20}$)}}}

\put(55,50){\line(0,-1){5}} \put(45,45){\line(1,0){20}}
\put(45,45){\vector(0,-1){5}}\put(65,45){\vector(0,-1){5}}
\put(35,35){\makebox(10,5)[c]{{\tiny
($\tilde{q}_{10}\odot\tilde{\sigma}_1$\hskip 4mm
   $\tilde{q}_{20}\odot\tilde{\sigma}_1$)}
}}\put(65,35){\makebox(10,5)[c]{{\tiny
($\tilde{q}_{10}\odot\tilde{\sigma}_2$\hskip 4mm
   $\tilde{q}_{20}\odot\tilde{\sigma}_2$)}}}
\put(40,35){\vector(0,-1){10}}

\put(15,30){\line(1,0){25}}

\put(15,30){\vector(0,-1){5}}

\put(70,35){\vector(0,-1){10}}

\put(70,30){\line(1,0){25}} \put(95,30){\vector(0,-1){5}}

 \put(5,20){\makebox(10,5)[c]{{\tiny
($\tilde{q}_{10}\odot\tilde{\sigma}_1^{2}$\hskip 4mm
$\tilde{q}_{20}\odot\tilde{\sigma}_1^{2}$)}
}}\put(34,20){\makebox(10,5)[c]{{\tiny
($\tilde{q}_{10}\odot\tilde{\sigma}_1\odot\tilde{\sigma}_2$\hskip
4mm $\tilde{q}_{20}\odot\tilde{\sigma}_1\odot\tilde{\sigma}_2$)}}}
\put(67,20){\makebox(10,5)[c]{{\tiny
($\tilde{q}_{10}\odot\tilde{\sigma}_2\odot\tilde{\sigma}_1$\hskip
4mm  $\tilde{q}_{20}\odot\tilde{\sigma}_2\odot\tilde{\sigma}_1$)}
}}\put(96,20){\makebox(10,5)[c]{{\tiny
($\tilde{q}_{10}\odot\tilde{\sigma}_2^{2}$\hskip 4mm
$\tilde{q}_{20}\odot\tilde{\sigma}_2^{2}$)}}}
\put(45,40){\makebox(5,5)[c]{{\tiny
$\tilde{\sigma}_1$}}}\put(60,40){\makebox(5,5)[c]{{\tiny
$\tilde{\sigma}_2$}}} \put(15,25){\makebox(5,5)[c]{{\tiny
$\tilde{\sigma}_1$}}}\put(35,25){\makebox(5,5)[c]{{\tiny
$\tilde{\sigma}_2$}}} \put(70,25){\makebox(5,5)[c]{{\tiny
$\tilde{\sigma}_1$}}}\put(90,25){\makebox(5,5)[c]{{\tiny
$\tilde{\sigma}_2$}}}

\put(5,17){\makebox(10,5)[c]{{\vdots}}}
\put(34,17){\makebox(10,5)[c]{{\vdots}}}
\put(66,17){\makebox(10,5)[c]{{\vdots}}}
 \put(96,17){\makebox(10,5)[c]{{\vdots}}}

\put(50,8){\makebox(20,5)[c]{{\footnotesize Fig. 5. The computing
tree deciding the all different pairs of fuzzy states reachable
from the initial} }} \put(25,5){\makebox(5,5)[c]{{\footnotesize
fuzzy state pair $(\tilde{q}_{10},\tilde{q}_{20})$.} }}
\end{picture}

\vskip -10mm

{\it Example 3.}  For max-min automata
$\tilde{G}=(\tilde{Q},\tilde{\Sigma},\tilde{\delta},\tilde{q}_{0})$
as Example 2, and
$\tilde{H}=(\tilde{Q}_{1},\tilde{\Sigma},\tilde{\gamma},\tilde{p}_{0})$,
where $\tilde{p}_{0}=[0.8\hskip 2mm 0.2]$,
$\tilde{\Sigma}=\{\tilde{\alpha}_{1},\tilde{\alpha}_{2}\}$ with
$\tilde{\alpha}_{1}=\left[\begin{array}{cc}
0.2&0.8\\
0.2&0.2
\end{array}
\right]$, $\tilde{\alpha}_{2}=\left[\begin{array}{cc}
0.2&0.2\\
0.8&0.5
\end{array}
\right]$, then from the above computing tree Fig. 5 we can derive
the computing tree Fig. 6 for deciding the set of all different
fuzzy state pairs reachable from $(\tilde{q}_{0},\tilde{p}_{0})$
as follows:

\setlength{\unitlength}{0.05in}

\begin{picture}(103,87)

\put(65,91){\makebox(5,5)[c]{{\tiny Begin}}}

\put(68,92){\vector(0,-1){2}}
 \put(72,86){\makebox(10,5)[c]{{\tiny ($\tilde{q}_0$,$\tilde{p}_0$)=([0.9 0.1],[0.8 0.2])}}}

\put(68,87){\line(0,-1){3}} \put(58,84){\line(1,0){20}}
\put(58,84){\vector(0,-1){5}}\put(78,84){\vector(0,-1){5}}
\put(53,75){\makebox(10,5)[c]{{\tiny ([0.4 0.8],[0.2 0.8])}
}}\put(73,75){\makebox(10,5)[c]{{\tiny ([0.4\hskip 1mm
0.2],[0.2\hskip 1mm 0.2])}}}

\put(58,76){\vector(0,-1){8}}\put(33,73){\line(1,0){25}}
\put(33,73){\vector(0,-1){5}}

\put(78,76){\vector(0,-1){8}}\put(78,73){\line(1,0){25}}
\put(103,73){\vector(0,-1){5}}
\put(28,64){\makebox(10,5)[c]{{\tiny ([0.4\hskip 1mm
0.4],[0.2\hskip 1mm 0.2])} }}\put(51,64){\makebox(10,5)[c]{{\tiny
([0.8\hskip 1mm 0.5],[0.8\hskip 1mm 0.5])}}}
\put(73,64){\makebox(10,5)[c]{{\tiny ([0.4\hskip 1mm
0.4],[0.2\hskip 1mm 0.2])} }}\put(97,64){\makebox(10,5)[c]{{\tiny
$\underline {([0.4\hskip 1mm 0.2],[0.2\hskip 1mm 0.2])}$}}}

\put(53,65){\line(0,-1){3}}       \put(43,62){\line(1,0){20}}
\put(43,62){\vector(0,-1){5}}\put(63,62){\vector(0,-1){5}}
\put(40,53){\makebox(8,5)[c]{{\tiny $\underline{([0.4\hskip 1mm
0.8],[0.2\hskip 1mm 0.8])}$} }}\put(59,53){\makebox(8,5)[c]{{\tiny
([0.5 0.5],[0.5 0.5])}}}

\put(28,65){\vector(0,-1){8}}\put(3,62){\line(1,0){25}}
\put(3,62){\vector(0,-1){5}}
\put(78,65){\vector(0,-1){8}}\put(78,62){\line(1,0){25}}
\put(103,62){\vector(0,-1){5}} \put(0,53){\makebox(8,5)[c]{{\tiny
$\underline {([0.4\hskip 1mm 0.4],[0.2\hskip 1mm 0.2]}$)}
}}\put(19,53){\makebox(8,5)[c]{{\tiny $\underline {([0.4\hskip 1mm
0.4],[0.2\hskip 1mm 0.2])}$}}}
\put(79,53){\makebox(8,5)[c]{{\tiny$\underline {([0.4\hskip 1mm
0.4],[0.2\hskip 1mm 0.2])}$}}}\put(98,53){\makebox(8,5)[c]{{\tiny
$\underline {([0.4\hskip 1mm 0.4],[0.2\hskip 1mm 0.2])}$}}}

\put(63,54){\line(0,-1){3}} \put(53,51){\line(1,0){20}}
\put(53,51){\vector(0,-1){5}}\put(73,51){\vector(0,-1){5}}
\put(45,41){\makebox(10,5)[c]{{\tiny ([0.4\hskip 1mm
0.4],[0.2\hskip 1mm 0.5])} }}\put(71,41){\makebox(10,5)[c]{{\tiny
$\underline {([0.5\hskip 1mm 0.5],[0.5\hskip 1mm 0.5])}$}}}

\put(48,42){\line(0,-1){3}} \put(38,39){\line(1,0){20}}
\put(38,39){\vector(0,-1){5}}\put(58,39){\vector(0,-1){5}}
\put(30,29){\makebox(10,5)[c]{{\tiny ([0.4\hskip 1mm
0.4],[0.2\hskip 1mm 0.2])} }}\put(56,29){\makebox(10,5)[c]{{\tiny
([0.4 \hskip 1mm0.4],[0.5\hskip 1mm 0.5])}}}

\put(28,30){\line(0,-1){3}} \put(18,27){\line(1,0){20}}
\put(18,27){\vector(0,-1){5}}\put(38,27){\vector(0,-1){5}}
\put(13,17){\makebox(10,5)[c]{{\tiny $\underline {([0.4\hskip 1mm
0.4],[0.2\hskip 1mm 0.2])}$}
}}\put(31,17){\makebox(10,5)[c]{{\tiny $\underline {([0.4\hskip
1mm 0.4],[0.2\hskip 1mm 0.2])}$}}}

\put(63,30){\line(0,-1){3}} \put(53,27){\line(1,0){20}}
\put(53,27){\vector(0,-1){5}}\put(73,27){\vector(0,-1){5}}
\put(50,17){\makebox(10,5)[c]{{\tiny ([0.4\hskip 1mm
0.4],[0.2\hskip 1mm 0.1])} }}\put(68,17){\makebox(10,5)[c]{{\tiny
$\underline {([0.4\hskip 1mm  0.4],[0.5\hskip 1mm 0.5])}$}}}

\put(53,18){\line(0,-1){3}} \put(43,15){\line(1,0){20}}
\put(43,15){\vector(0,-1){5}}\put(63,15){\vector(0,-1){5}}
\put(40,5){\makebox(10,5)[c]{{\tiny $\underline {([0.4\hskip 1mm
0.4],[0.2\hskip 1mm 0.2])}$} }}\put(58,5){\makebox(10,5)[c]{{\tiny
$\underline {([0.4\hskip 1mm  0.4],[0.2\hskip 1mm 0.2])}$}}}

\put(55,79){\makebox(10,5)[c]{{\tiny $\tilde{\alpha}_1$}}}
\put(71,79){\makebox(10,5)[c]{{\tiny $\tilde{\alpha}_2$}}}
\put(30,68){\makebox(10,5)[c]{{\tiny $\tilde{\alpha}_1$}}}
\put(51,68){\makebox(10,5)[c]{{\tiny $\tilde{\alpha}_2$}}}
\put(75,68){\makebox(10,5)[c]{{\tiny $\tilde{\alpha}_1$}}}
\put(96,68){\makebox(10,5)[c]{{\tiny $\tilde{\alpha}_2$}}}

\put(0,57){\makebox(10,5)[c]{{\tiny $\tilde{\alpha}_1$}}}
\put(21,57){\makebox(10,5)[c]{{\tiny $\tilde{\alpha}_2$}}}
\put(75,57){\makebox(10,5)[c]{{\tiny $\tilde{\alpha}_1$}}}
\put(96,57){\makebox(10,5)[c]{{\tiny $\tilde{\alpha}_2$}}}
\put(50,46){\makebox(10,5)[c]{{\tiny $\tilde{\alpha}_1$}}}
\put(66,46){\makebox(10,5)[c]{{\tiny $\tilde{\alpha}_2$}}}
\put(40,57){\makebox(10,5)[c]{{\tiny $\tilde{\alpha}_1$}}}
\put(56,57){\makebox(10,5)[c]{{\tiny $\tilde{\alpha}_2$}}}
\put(35,34){\makebox(10,5)[c]{{\tiny $\tilde{\alpha}_1$}}}
\put(51,34){\makebox(10,5)[c]{{\tiny $\tilde{\alpha}_2$}}}
\put(15,22){\makebox(10,5)[c]{{\tiny $\tilde{\alpha}_1$}}}
\put(31,22){\makebox(10,5)[c]{{\tiny $\tilde{\alpha}_2$}}}
\put(50,22){\makebox(10,5)[c]{{\tiny $\tilde{\alpha}_1$}}}
\put(66,22){\makebox(10,5)[c]{{\tiny $\tilde{\alpha}_2$}}}
\put(40,10){\makebox(10,5)[c]{{\tiny $\tilde{\alpha}_1$}}}
\put(56,10){\makebox(10,5)[c]{{\tiny $\tilde{\alpha}_2$}}}
\put(50,1){\makebox(10,5)[c]{{\footnotesize Fig. 6. The computing
tree deciding the all different pairs of fuzzy states reachable
from the initial fuzzy}}}
\put(24,-2){\makebox(5,5)[c]{{\footnotesize state pair
($\tilde{q}_0$,$\tilde{p}_0$)=([0.9 0.1],[0.8 0.2]).}}}

\end{picture}

\noindent On the basis of the above discussion, if $pr(\tilde{K})$
is generated by a max-min automaton
$\tilde{H}=(\tilde{Q}_{1},\tilde{\Sigma},\tilde{\gamma},\tilde{p}_{0})$,
we can test the fuzzy controllability condition Eq. (9) within
finite steps by means of  TABLE II, where

$L(\tilde{G},\tilde{H},\tilde{s},\tilde{\sigma})=\min\left\{
pr(\tilde{K})(\tilde{s}),\tilde{\Sigma}_{uc}(\tilde{\sigma}),
{\cal L}_{\tilde{G}}( \tilde{s}\tilde{\sigma} )\right\}$,

$ pr(\tilde{K})(\tilde{s})={\cal L}_{\tilde{H}}(\tilde{s})=[\tilde{p}_{0}\odot\tilde{s}]$,

${\cal L}_{\tilde{G}}( \tilde{s}\tilde{\sigma} )=[\tilde{q}_{0}\odot\tilde{s}\odot\tilde{\sigma}]$.\\
Here the set of all different fuzzy state pairs is assumed to be
$$\{(\tilde{q}_{0}\odot\tilde{s},\tilde{p}_{0}\odot\tilde{s}):
\tilde{s}_{1},\tilde{s}_{2},\ldots,\tilde{s}_{m}\in\tilde{\Sigma}^{*}\}.$$
In TABLE II, $L(\tilde{G},\tilde{H},\tilde{s},\tilde{\sigma})\leq
[\tilde{p}_0\odot\tilde{s}\odot\tilde{\sigma}]$ is exactly the
fuzzy controllability condition Eq. (9); if all elements in the
rightmost column are ``T", then the fuzzy controllability
condition Eq. (9) holds true; otherwise, it does not hold.
\newpage

\setlength{\unitlength}{0.05in}

\begin{picture}(50,5)

\put(49,-2){\makebox(10,5)[c]{{\tiny TABLE  II. Testing the fuzzy
controllability condition Eq. (9) in terms of whether or not the
all elements in} }}

\put (26,-5){\makebox(10,5)[c]{{\tiny  the rightmost column are
``T (True) ".} }}

\end{picture}

\begin{center}
{\tiny
 \hskip -3mm
\begin {tabular}{|c|c|c|c|c|c|c|c|}
\hline
$\tilde{s}$&$\tilde{\sigma}$&[$\tilde{p}_0\odot\tilde{s}$]&[$\tilde{q}_0\odot\tilde{s}\odot\tilde{\sigma}$]&$\tilde{\Sigma}_{uc}(\tilde{\sigma})$&
$L(\tilde{G},\tilde{H},\tilde{s},\tilde{\sigma}$)&[$\tilde{p}_0\odot\tilde{s}\odot\tilde{\sigma}$]&
$L(\tilde{G},\tilde{H},\tilde{s},\tilde{\sigma}$)$\leq$[$\tilde{p}_0\odot\tilde{s}\odot\tilde{\sigma}$]\\
\hline
$\tilde{s}_1$&$\tilde{\sigma}_1$&&&&&& T (True) or F (False)\\
\cline{2-8}
  &$\vdots$&&&&&&T (True) or F (False)\\
\cline{2-8} &$\tilde{\sigma}_n$&&&&&&T (True) or F (False)\\
\hline
$\tilde{s}_2$&$\tilde{\sigma}_1$&&&&&&T (True) or F (False)\\
\cline{2-8}
  &$\vdots$&&&&&&T (True) or F (False)\\
\cline{2-8} &$\tilde{\sigma}_n$&&&&&&T (True) or F (False)\\
\hline $\vdots$&$\vdots$&&&&&&T (True) or F (False)\\
\hline
$\tilde{s}_m$&$\tilde{\sigma}_1$&&&&&&T (True) or F (False)\\
\cline{2-8}
  &$\vdots$&&&&&&T (True) or F (False)\\
\cline{2-8} &$\tilde{\sigma}_n$&&&&&&T (True) or F (False)\\
\hline

\end{tabular}}

\end{center}

{\it Example 4.} Let fuzzy DES be modeled by max-min automaton
$\tilde{G}$ in Example 2, and $pr(\tilde{K})$ is generated by
$\tilde{H}$ in Example 3. By virtue of Fig. 6 we know that the all
different pairs of fuzzy states reachable from
$(\tilde{q}_{0},\tilde{p}_{0})$ are listed in TABLE III.

\begin{picture}(50,5)

\put(39,0){\makebox(10,5)[c]{{\tiny TABLE  III. The computing tree
visualized in Fig. 6 shows the all different pairs of  fuzzy
states}}}

\put(19,-2){\makebox(10,5)[c]{{\tiny   reachable from
$(\tilde{q}_{0},\tilde{p}_{0})$.}}}
\end{picture}

\begin {tabular}{|c|c|c|c|}
\hline $\tilde{s}$&($\tilde{q}_0\odot\tilde{s}$,\hskip 2mm
$\tilde{p}_0\odot\tilde{s}$)&$\tilde{s}$&($\tilde{q}_0\odot\tilde{s}$,\hskip
2mm $\tilde{p}_0\odot\tilde{s}$)\\
\hline $\tilde{\epsilon}$&([0.9 \hskip 1mm 0.1], [0.8 \hskip 1mm
0.2])&$\tilde{\alpha}_1\tilde{\alpha}_2$&([0.8 \hskip 1mm 0.5],
[0.8
\hskip 1mm 0.5])\\
\hline $\tilde{\alpha}_1$&([0.4 \hskip 1mm 0.8], [0.2 \hskip 1mm
0.8])&$\tilde{\alpha}_1\tilde{\alpha}_2\tilde{\alpha}_2$&([0.5
\hskip 1mm 0.5], [0.5 \hskip 1mm  0.5])\\
\hline
 $\tilde{\alpha}_2$&([0.4 \hskip
1mm 0.2], [0.2 \hskip 1mm
0.2])&$\tilde{\alpha}_1\tilde{\alpha}_2\tilde{\alpha}_2\tilde{\alpha}_1$&([0.4
\hskip 1mm 0.4], [0.5 \hskip 1mm  0.5])\\
\hline

$\tilde{\alpha}_1\tilde{\alpha}_1$&([0.4 \hskip 1mm 0.4], [0.2
\hskip 1mm  0.2])&$\tilde{\alpha}_1\tilde{\alpha}_2\tilde{\alpha}_2\tilde{\alpha}_1\tilde{\alpha}_2$&([0.4 \hskip 1mm 0.4], [0.5 \hskip 1mm 0.5])\\
\hline
\end{tabular}

\vskip 2mm \noindent According to TABLE II, we can check the fuzzy
controllability condition Eq. (9) of fuzzy DES in Example 4 via
TABLE IV as follows  (we only write the cases of
$\tilde{s}=\epsilon$ and $\tilde{\alpha}_{1}$, since an ``F" has
been found up to now). \vskip -5mm
\begin{picture}(50,5)

\put(43,-4){\makebox(10,5)[c]{{\tiny TABLE  IV. Testing whether or
not the fuzzy controllability condition Eq. (9) of fuzzy DES in
Example 4 holds.}}}

\end{picture}

\begin{center}
{\footnotesize
 \hskip -3mm
\begin {tabular}{|c|c|c|c|c|c|c|c|}\hline
$\tilde{s}$&$\tilde{\sigma}$&[$\tilde{p}_0\odot\tilde{s}$]&[$\tilde{q}_0\odot\tilde{s}\odot\tilde{\sigma}$]&$\tilde{\Sigma}_{uc}(\tilde{\sigma})$&
$L(\tilde{G},\tilde{H},\tilde{s},\tilde{\sigma}$)&[$\tilde{p}_0\odot\tilde{s}\odot\tilde{\sigma}$]&
$L(\tilde{G},\tilde{H},\tilde{s},\tilde{\sigma}$)$\leq$[$\tilde{p}_0\odot\tilde{s}\odot\tilde{\sigma}$]\\
\hline
$\tilde{\epsilon}$&$\tilde{\alpha}_1$&0.8&0.8&0.7&0.7&0.8&T\\

\cline{2-8} &$\tilde{\alpha}_2$&0.8&0.4&0.2&0.2&0.2&T\\
\hline
$\tilde{\alpha}_1$&$\tilde{\alpha}_1$&0.8&0.4&0.7&0.4&0.2&F\\

\cline{2-8} &$\tilde{\alpha}_2$&0.8&0.8&0.2&0.2&0.8&T\\
\hline

\end{tabular}
}
\end{center}

\noindent Therefore by means of TABLE IV we see that Eq. (9) does
not hold for $\tilde{s}=\tilde{\alpha}_{1}$ and
$\tilde{\sigma}=\tilde{\alpha}_{1}$.

\subsection*{{\it C. A Primary Application to Medicine: A Formal Example}}

We further state some applicable background of fuzzy DESs. Fuzzy
control and fuzzy systems in biomedical engineering have been
significantly studied (for example, see [6]), in which fuzzy drug
delivery system for real-time control of mean arterial pressure
(MAP), cardiac output (CO), and mean pulmonary arterial pressure
(MPAP) in patients is one of the main concerns. The heart
patient's status may be represented via the degrees of the three
hemodynamic variables, i.e., MAP, CO, and MPAP, and they may be
low, high, or normal. These drugs such as dopamine (DPM), sodium
nitroprusside (SNP), nitroglycerin (NTG), and phenylephrine (PNP)
are appropriately used to regulate the degrees of MAP, CO, and
MPAP. It may be logically modeled via fuzzy DESs with supervisory
control, in which the uses of the dosages of DPM, SNP, NTG, and
PNP may be thought of as some fuzzy events. But here we consider
the patient's condition roughly to be three cases, i.e., ``poor",
``fair", and ``excellent", and, as a first step, apply the above
results regarding supervised control of fuzzy DESs to control the
three states. To this end, we further describe the following
examples, which may be viewed as an applicable background of
supervisory control of fuzzy DESs.

\setlength{\unitlength}{0.05in}

\begin{picture}(80,22)

\put(25,-5){\circle{4}} \put(10,15){\circle{4}}
\put(40,15){\circle{4}}

\put(22.5,-4){\vector(-3,4){12.5}}
\put(11.5,13.5){\vector(3,-4){12.5}}
\put(26,-3){\vector(3,4){12.5}} \put(40,13){\vector(-3,-4){12.5}}
\put(12,14.5){\vector(1,0){26}} \put(38,16){\vector(-1,0){26.5}}
\put(25,-9){\makebox(0.05,0.05)[c]{$p_{2}$}}
\put(5,15){\makebox(0.05,0.05)[c]{$p_{1}$}}
\put(45,15){\makebox(0.05,0.05)[c]{$p_{3}$}}
\put(25,13){\makebox(0.05,0.05)[c]{$\alpha_{3}$}}
\put(25,17.5){\makebox(0.05,0.05)[c]{$\beta_{3}$}}
\put(19.5,6){\makebox(0.05,0.05)[c]{$\alpha_{1}$}}
\put(14.5,4){\makebox(0.05,0.05)[c]{$\beta_{2}$}}
\put(30.5,6){\makebox(0.05,0.05)[c]{$\alpha_{2}$}}
\put(35.5,4){\makebox(0.05,0.05)[c]{$\beta_{1}$}}
\put(23,-13){\makebox(0.5,0.05)[c]{{\tiny Fig. 7. Finite automaton
modelling patient's heart condition.} }}

\put(75,-5){\circle{4}} \put(60,15){\circle{4}}
\put(90,15){\circle{4}}

\put(75,-5){\circle{3}} \put(90,15){\circle{3}}

\put(61.5,13.5){\vector(3,-4){12.5}}

\put(76,-3){\vector(3,4){12.5}}

\put(62,14.5){\vector(1,0){26}}
\put(75,-9){\makebox(0.05,0.05)[c]{$p_{2}$}}
\put(55,15){\makebox(0.05,0.05)[c]{$p_{1}$}}
\put(95,15){\makebox(0.05,0.05)[c]{$p_{3}$}}

\put(75,13){\makebox(0.05,0.05)[c]{$\alpha_{3}$}}

\put(69.5,6){\makebox(0.05,0.05)[c]{$\alpha_{1}$}}

\put(80.5,6){\makebox(0.05,0.05)[c]{$\alpha_{2}$}}

\put(80,-13){\makebox(0.5,0.05)[c]{{\tiny Fig. 8. Finite automaton
modelling the given specifications.} }}

\end{picture}

\vskip 15mm

{\it Example 5.} Let us use automaton $G=(Q,\Sigma,\delta,p_{1})$
to model a patient's heart condition, where
$Q=\{p_{1},p_{2},p_{3}\}$,
$\Sigma=\{\alpha_{1},\alpha_{2},\alpha_{3},\beta_{1},\beta_{2},\beta_{3}\}$,
and $\delta$ is visualized by Fig. 7. Here $p_{1}=P$, $p_{2}=F$,
and $p_{3}=E$ represent ``poor", ``fair", and ``excellent",
respectively. A patient's initial condition may be ``poor" and
should become ``fair" and even ``excellent" after certain
treatment. When patient's health gets ``fair", we naturally hope
it to be better and better, say ``excellent", instead of
deteriorating, i.e., recurring to ``poor". Analogously, if the
patient's condition has been ``excellent", it is desired to keep
the good health and thus a supervisor is necessary to disable the
events $\beta_{1}$, $\beta_{2}$, and $\beta_{3}$ in case they are
controllable. Let us assume
$\Sigma_{uc}=\{\alpha_{1},\alpha_{2},\alpha_{3}\}$,
$\Sigma_{c}=\{\beta_{1},\beta_{2},\beta_{3}\}$, and the given set
of specifications
$K=\{\alpha_{1},\alpha_{1}\alpha_{2},\alpha_{3}\}$ as desired.
Actually, $K$ is marked by automaton
$H=(Q,\Sigma_{uc},\delta_{1},p_{1},Q_{m})$ depicted by Fig. 8,
where $Q_{m}=\{p_{2},p_{3}\}$. Clearly,
$pr(K)=\{\epsilon,\alpha_{1},\alpha_{1}\alpha_{2},\alpha_{3}\}$ is
generated by $H$. In terms of the method described above for
checking the controllability condition Eq. (9), it is readily seen
that $K$ is controllable with respect to $L(G)$ and $\Sigma_{uc}$,
that is, $pr(K)\Sigma_{uc}\cap L(G)\subseteq pr(K)$ holds, so by
virtue of crisp controllable theorem indicated above, there is a
supervisor $S:\Sigma^{*}\rightarrow {\cal P}(\Sigma)$ such that
$L(S/G)=pr(K)$, where $S$ is defined as: For any $s\in pr(K)$,
$S(s)=\Gamma_{H}(\delta_{1}(p_{1},s))$. Here symbol
$\Gamma_{H}(q)$ denotes the set of active events in the current
state $q$. More explicitly,
$S(\epsilon)=\{\alpha_{1},\alpha_{3}\}$,
$S(\alpha_{1})=\{\alpha_{2}\}$,
$S(\alpha_{1}\alpha_{2})=\emptyset$, $S(\alpha_{3})=\emptyset$.

In real-life situation, a patient's condition can simultaneously
belong to ``excellent", ``fair", and even ``poor" with respective
memberships; also, an event occurring (i.e., a treatment) may lead
a state to multi-states with respective degrees. Therefore, a
patient's conditions and their changes after certain treatments
should be modeled by a max-min automaton
$\tilde{G}=(\tilde{Q},\tilde{\Sigma},\tilde{\delta},\tilde{q}_{0})$,
and the fuzzy events $\tilde{\alpha}_{i}$, $\tilde{\beta}_{i}$
$(i=1,2)$ may be evaluated as follows by means of some diagnosis
together with medical theory and experience (for example, by
virtue of an intelligent system designed for automating drug
delivery in [6, page 323], which  is described by Fig. 9, and to a
certain extent, similar to the process of supervisory control of
fuzzy DESs).

\begin{picture}(50,40)

\put(5,8){\line(1,0){22}} \put(5,8){\line(0,1){6}}

\put(27,8){\line(0,1){6}}\put(5,14){\line(1,0){22}}
 \put(5,11){\makebox(22,2.5)[c]{{\tiny
Fuzzy Decision-Making}}} \put(5,8){\makebox(22,2.5)[c]{{\tiny
Module (FDMM)}}} \put(27,11){\vector(1,0){10}}

\put(37,8){\line(1,0){22}} \put(37,8){\line(0,1){6}}

\put(59,8){\line(0,1){6}}\put(37,14){\line(1,0){22}}
\put(37,11){\makebox(22,2.5)[c]{{\tiny Fuzzy Control}}}
\put(37,8){\makebox(22,2.5)[c]{{\tiny  Module (FCM)}}}
 \put(59,11){\vector(1,0){10}}

\put(69,8){\line(1,0){22}} \put(69,8){\line(0,1){6}}
\put(91,8){\line(0,1){6}}\put(69,14){\line(1,0){22}}
\put(69,11){\makebox(22,2.5)[c]{{\tiny Therapeutic Assessment}}}
\put(69,8){\makebox(22,2.5)[c]{{\tiny Module (TAM)}}}
 \put(15,34){\vector(0,-1){20}}
 \put(15,34){\line(1,0){22}}

\put(37,31){\line(1,0){22}} \put(37,31){\line(0,1){6}}
\put(59,31){\line(0,1){6}} \put(37,37){\line(1,0){22}}
\put(37,34){\makebox(22,2.5)[c]{{\tiny Plant}}}
\put(37,31){\makebox(22,2.5)[c]{{\tiny (Patient)}}}
 \put(80,34){\vector(-1,0){21}}\put(80,34){\line(0,-1){20}}
 \put(48,31){\vector(0,-1){17}}

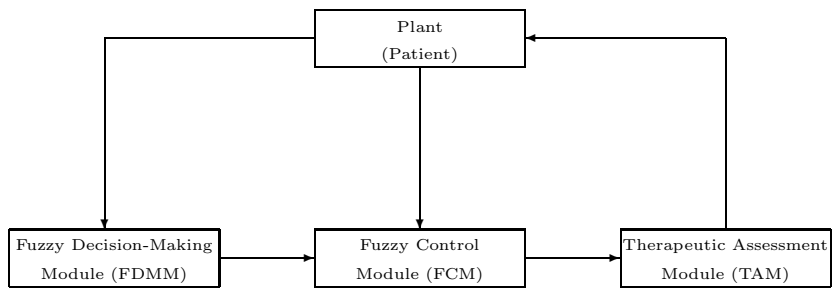
\put(45,0){\makebox(10,5)[c]{{\tiny Fig.  9. An intelligent system
may be used to evaluate fuzzy events, where FDMM may be utilized
to evaluate  }}}

\put(48,-2){\makebox(10,5)[c]{{\tiny the status of the patient,
FCM may be devised to decide the fuzzy events to be carried out,
and TAM }}} \put(7,-4){\makebox(10,5)[l]{{\tiny  is used to
evaluate the feasibility of implemented fuzzy events.}}}
\end{picture}

\[
\tilde{\alpha}_{1}=
\left[\begin{array}{lcc} 0.4&0.9&0.4\\
0&0.4&0.4\\
0&0&0.4
\end{array}
\right], \hskip 3mm
 \tilde{\alpha}_{2}=
\left[\begin{array}{lcc} 0.4&0.4&0.4\\
0&0.4&0.9\\
0&0&0.4
\end{array}
\right],\hskip 3mm
\tilde{\alpha}_{3}=
\left[\begin{array}{lcc} 0.4&0.4&0.9\\
0&0.4&0.4\\
0&0&0.4
\end{array}
\right],
\]
\[
\tilde{\beta}_{1}=
\left[\begin{array}{lcc} 0.4&0&0\\
0.4&0.4&0\\
0.4&0.9&0.4
\end{array}
\right], \hskip 3mm
 \tilde{\beta}_{2}=
\left[\begin{array}{lcc} 0.4&0&0\\
0.9&0.4&0\\
0.4&0.4&0.4
\end{array}
\right], \hskip 3mm
\tilde{\beta}_{3}=
\left[\begin{array}{lcc} 0.4&0&0\\
0.4&0.4&0\\
0.9&0.4&0.4
\end{array}
\right];
\]
$\tilde{\delta}$ is defined as before; $\tilde{q}_{0}=[0.9\hskip
2mm 0.1\hskip 2mm 0]$. As analyzed above, we hope the patient's
condition to be better and better. To this end, we specify a fuzzy
set of control specifications $ \tilde{K} \in {\cal
F}(\tilde{\Sigma}^{*})$ that are desired, and show whether there
exists a supervisor that can disable some fuzzy events with
respective degrees such that the restricted behavior of the
supervised fuzzy system satisfies those given specifications.

As usual, let $pr(\tilde{K})$ be generated by a max-min automaton
$\tilde{H}=(\tilde{Q},\tilde{\Sigma},\tilde{\delta}_{1},\tilde{q}_{0})$,
that is assumed to be derived from an intelligent system mentioned
above,  where $\tilde{q}_{0}=[0.9\hskip 2mm 0.1\hskip 2mm 0]$;
$\tilde{\Sigma}= \{
\tilde{\alpha}_{1},\tilde{\alpha}_{2},\tilde{\alpha}_{3},
\tilde{\beta}_{1}, \tilde{\beta}_{2}, \tilde{\beta}_{3}\}$ where
$\tilde{\alpha}_{1}$, $\tilde{\alpha}_{2}$, $\tilde{\alpha}_{3}$
are the same as those in $\tilde{G}$, but $\tilde{\beta}_{1}$,
$\tilde{\beta}_{2}$, $\tilde{\beta}_{3}$ are changed as follows:
\[
\tilde{\beta}_{1} =
\left[\begin{array}{lcc} 0.2&0&0\\
0.2&0.2&0\\
0.2&0.9&0.2
\end{array}
\right], \hskip 3mm
 \tilde{\beta}_{2}=
\left[\begin{array}{lcc} 0.2&0&0\\
0.9&0.2&0\\
0.2&0.2&0.2
\end{array}
\right], \hskip 3mm
\tilde{\beta}_{3}=
\left[\begin{array}{lcc} 0.2&0&0\\
0.2&0.2&0\\
0.9&0.2&0.2
\end{array}
\right];
\]
$\tilde{\delta}_{1}$ is naturally defined in terms of these given
fuzzy events. Clearly, $pr(\tilde{K})= {\cal L}_{\tilde{H}}
\subseteq {\cal L}_{\tilde{G}}$ since each element in every fuzzy
event of $\tilde{H}$ is not bigger than the corresponding element
in the same fuzzy event of $\tilde{G}$. By means of the computing
tree described by Fig. 6, we can obtain that there are only twelve
different pairs of fuzzy states reachable from
$(\tilde{q}_{0},\tilde{q}_{0})$, that are listed in TABLE V.

\begin{picture}(50,5)

\put(1,-2){\makebox(10,5)[l]{{\tiny TABLE  V. The all different
pairs of fuzzy states reachable from
$(\tilde{q}_{0},\tilde{q}_{0})$.} }}

\end{picture}

{\footnotesize
\begin {tabular}{|c|c|c|c|}
\hline $\tilde{s}$&($\tilde{q}_0\odot\tilde{s}$,\hskip 2mm
$\tilde{q}_0\odot\tilde{s}$)&$\tilde{s}$&($\tilde{q}_0\odot\tilde{s}$,\hskip
2mm  $\tilde{q}_0\odot\tilde{s}$)\\
\hline
$\tilde{\epsilon}$&([0.9 \hskip 1mm 0.1 \hskip 1mm 0], [0.9
\hskip 1mm  0.1 \hskip 1mm
0])&$\tilde{\alpha}_1\tilde{\beta}_2$&([0.9 \hskip 1mm 0.4 \hskip
1mm 0.4], [0.9 \hskip 1mm 0.2 \hskip 1mm 0.2])\\
\hline $\tilde{\alpha}_1$&([0.4 \hskip 1mm 0.9 \hskip 1mm 0.4],
[0.4 \hskip 1mm 0.9 \hskip 1mm 0.4])&$\tilde{\alpha}_2\tilde{\beta}_2$&([0.4 \hskip 1mm 0.4 \hskip 1mm 0.4], [0.4 \hskip 1mm 0.2 \hskip 1mm 0.2])\\
\hline $\tilde{\alpha}_2$&([0.4 \hskip 1mm 0.4 \hskip 1mm 0.4],
[0.4 \hskip 1mm 0.4 \hskip 1mm
0.4])&$\tilde{\alpha}_3\tilde{\beta}_1$&([0.4 \hskip 1mm 0.9 \hskip 1mm 0.4], [0.2 \hskip 1mm 0.9 \hskip 1mm 0.2])\\

\hline
 $\tilde{\alpha}_3$&([0.4 \hskip 1mm 0.4 \hskip 1mm 0.9], [0.4 \hskip 1mm 0.4 \hskip 1mm 0.9])
 &$\tilde{\alpha}_3\tilde{\beta}_1\tilde{\alpha}_1$&([0.4 \hskip
1mm 0.4 \hskip 1mm 0.4], [0.2 \hskip 1mm 0.4 \hskip 1mm
0.4])\\
\hline $\tilde{\beta}_1$&([0.4 \hskip 1mm 0.1 \hskip 1mm 0], [0.2
\hskip 1mm 0.1 \hskip 1mm
0])&$\tilde{\alpha}_3\tilde{\beta}_1\tilde{\alpha}_2$&([0.4 \hskip 1mm 0.4 \hskip 1mm 0.9], [0.2 \hskip 1mm 0.4 \hskip 1mm 0.9])\\
\hline $\tilde{\alpha}_1\tilde{\beta}_1$&([0.4 \hskip 1mm 0.4
\hskip 1mm 0.4], [0.2 \hskip 1mm 0.2 \hskip 1mm
0.2])&$\tilde{\alpha}_3\tilde{\beta}_1\tilde{\alpha}_1\tilde{\beta}_1$&([0.4 \hskip 1mm 0.4 \hskip 1mm 0.4], [0.2 \hskip 1mm 0.4 \hskip 1mm 0.2])\\

\hline

\end{tabular}
}

Suppose that the fuzzy set of uncontrollable events $
\tilde{\Sigma}_{uc} $ is also evaluated in terms of medical theory
and experience as: $ \tilde{\Sigma}_{uc}(
\tilde{\alpha_{1}})=0.8$;
$\tilde{\Sigma}_{uc}(\tilde{\alpha_{2}})=0.75$;
$\tilde{\Sigma}_{uc}(\tilde{\alpha_{3}})= 0.7$; $
\tilde{\Sigma}_{uc}(\tilde{\beta_{1}}) = 0.2$;
$\tilde{\Sigma}_{uc}(\tilde{\beta_{2}})=0.25$;
$\tilde{\Sigma}_{uc}(\tilde{\beta_{3}})= 0.3$; and
$\tilde{\Sigma}_{c} $ is naturally decided due to $
\tilde{\Sigma}_{uc} (\tilde{\sigma})+ \tilde{\Sigma}_{c}
(\tilde{\sigma})=1$ for any fuzzy event $\tilde{\sigma}\in \{
\tilde{\alpha_{1}},\tilde{\alpha_{2}},\tilde{\alpha_{3}},
\tilde{\beta_{1}}, \tilde{\beta_{2}}, \tilde{\beta_{3}}\}$. Then
by virtue of TABLE I we can test that the fuzzy controllability
condition Eq. (9) does not hold for $\tilde{s}=\epsilon$,
$\tilde{\sigma}=\tilde{\beta}_{2}$ or $\tilde{\beta}_{3}$. If
$\tilde{\Sigma}_{uc}\leq 0.2$ for any $\tilde{\sigma}\in
\tilde{\Sigma}$, then it can be checked by using TABLE I that Eq.
(9) holds true, and in terms of Theorem 1 we know that the
supervisor $\tilde{S}$ yielding $pr(\tilde{K})$ therefore exists.
In practice, the degree of each fuzzy event being controlled by
the supervisor is specified according to Eq. (10), and the
therapeutic process is then implemented under the control of the
supervisor, which results in the desired specification
$pr(\tilde{K})$.

\subsection*{{\it C. Nonblocking Controllability Theorem for Fuzzy DESs}}

In supervisory control of crisp DESs, nonblockingness is usually
required and it means that the production sequence is completed
without deadlock, which has been applied to manufacturing to avoid
deadlock. In reality, analogous restriction is sometimes imposed
on fuzzy DESs. More exactly, the specifications on that controlled
system are given as a sublanguage of $ {\cal L}_{\tilde{G},m}$,
and the supervisor $\tilde{S}$ is {\it nonblocking}, that is,
\[
pr( {\cal L}_{ \tilde{S}/\tilde{G} ,m}) = {\cal L}_{\tilde{S}/\tilde{G}};
\]
otherwise, $\tilde{S}$ is {\it blocking}. Intuitively, $\tilde{S}$
being nonblocking means that for any string of fuzzy events
$\tilde{s}$, the degree to which $\tilde{s}$ belongs to the fuzzy
language generated by the supervised fuzzy system
$\tilde{S}/\tilde{G} $ equals the membership degree of $\tilde{s}$
belonging to the prefix-closure of the fuzzy language marked by
the supervised fuzzy system $ \tilde{S}/\tilde{G} $. In Example 5,
if the state ``excellent" is indicated as a marked state,  then
the nonblockingness renders that any partial therapeutic process
(say $\tilde{\alpha}_{1}$) being implemented under the supervisor
$\tilde{S}$ has the possibility not more than that of the whole
expected therapeutic strategy (say
$\tilde{\alpha}_{1}\tilde{\alpha}_{2}$) supervised by $\tilde{S}$
being realized, which also conforms to the crisp situation. We
provide another numeric example to serve as further illustration.

{\it Example 6.} Let set of fuzzy events
$\tilde{\Sigma}=\{\tilde{a}, \tilde{b}, \tilde{c}\}$. Assume that
the fuzzy language ${\cal L}_{\tilde{G}}$ and ${\cal
L}_{\tilde{G},m}$ generated and marked by a fuzzy finite automaton
(max-min or max-product) $\tilde{G}$ are respectively as follows:
${\cal L}_{\tilde{G}}(\epsilon)= 1$, $ {\cal
L}_{\tilde{G}}(\tilde{a})= {\cal L}_{\tilde{G}}
(\tilde{a}\tilde{b})= {\cal
L}_{\tilde{G}}(\tilde{a}\tilde{b}\tilde{c})= 0.8$, and ${\cal
L}_{\tilde{G}}(\tilde{s})=0$ for the other $s\in
\tilde{\Sigma}^{*}$; ${\cal L}_{\tilde{G},m}(\epsilon)= 1$, ${\cal
L}_{\tilde{G},m}(\tilde{a}\tilde{b})= 0.8$, and ${\cal
L}_{\tilde{G},m}(\tilde{s})=0$ for the other $s\in
\tilde{\Sigma}^{*}$. If $ \tilde{\Sigma}_{uc}(\tilde{a})
=\tilde{\Sigma}_{uc}(\tilde{b})=0.8$ and
$\tilde{\Sigma}_{uc}(\tilde{c})=0$, we define a supervisor
$\tilde{S}:\tilde{\Sigma}^{*}\rightarrow {\cal F}(\tilde{\Sigma})$
as follows: $\tilde{S}(\epsilon)(\tilde{a})=1$, $\tilde{S}(
\tilde{a})(\tilde{b})=0.8$, $\tilde{S}( \tilde{a}\tilde{b})
(\tilde{c})=0$, and also 0 for the other cases. Then it is easy to
check that $\tilde{S}$ satisfies the fuzzy admissibility condition
described by Eq. (6), and
\[
{\cal L}_{\tilde{S}/\tilde{G} }(\tilde{a})=pr( {\cal L}_{ \tilde{S}/\tilde{G} ,m})(\tilde{a})=0.8,
\]
\[
{\cal L}_{\tilde{S}/\tilde{G} }(\tilde{a}\tilde{b})= pr( {\cal L}_{ \tilde{S}/\tilde{G} ,m})(\tilde{a}\tilde{b})=0.8,
\]
and ${\cal L}_{\tilde{S}/\tilde{G} }(\tilde{s})=pr( {\cal L}_{
\tilde{S}/\tilde{G} ,m})(\tilde{s})=0$ for the other $\tilde{s}\in
\tilde{\Sigma}^{*}$. Therefore, in this case, $\tilde{S}$ is {\it
nonblocking}. However, if $ \tilde{\Sigma}_{uc}(\tilde{c}) >0$,
and still $ \tilde{\Sigma}_{uc}(\tilde{a})
=\tilde{\Sigma}_{uc}(\tilde{b})=0.8$, then from the fuzzy
admissibility condition it follows that
\[
\tilde{S}(\tilde{a}\tilde{b})(\tilde{c})\geq \min\{ \tilde{ \Sigma}_{uc}(\tilde{c}), {\cal L}_{\tilde{G}}(\tilde{a}\tilde{b}\tilde{c})\}
=\min\{ \tilde{ \Sigma}_{uc}(\tilde{c}), 0.8\}>0.
\]
and thus ${\cal L}_{\tilde{S}/\tilde{G}
}(\tilde{a}\tilde{b}\tilde{c})=\min\{\Sigma_{uc}(\tilde{c}),0.8\}$,
but $pr( {\cal L}_{ \tilde{S}/\tilde{G}
,m})(\tilde{a}\tilde{b}\tilde{c})=0$. Therefore in this case $pr(
{\cal L}_{ \tilde{S}/\tilde{G} ,m}) \not= {\cal
L}_{\tilde{S}/\tilde{G}}$, which shows that the supervisor
$\tilde{S}$ is {\it blocking}.\hskip 5mm $\Box$

In this regard, we have the nonblocking controllability theorem of
fuzzy DESs.

{\it Theorem 2.} Let a fuzzy DES be modeled by fuzzy finite
automaton (max-min or max-product)
$\tilde{G}=(\tilde{Q},\tilde{\Sigma},\tilde{\delta},\tilde{q}_{0},\tilde{Q}_{m})$,
and let $\tilde{\Sigma}_{uc} \in {\cal F}(\tilde{\Sigma})$ be the
fuzzy uncontrollable subset of $\tilde{\Sigma}$. Suppose fuzzy
language $\tilde{K}\subseteq {\cal L}_{\tilde{G},m}$ satisfying
$\tilde{K}(\epsilon)=1$ and $pr(\tilde{K})\subseteq {\cal
L}_{\tilde{G},m}$. Then there exists a nonblocking supervisor
$\tilde{S}$ for $\tilde{G}$ such that $\tilde{S}$ satisfies the
fuzzy admissibility condition Eq. (6), and
\[
{\cal L}_{\tilde{S}/\tilde{G},m}= \tilde{K}    \hskip 2mm {\rm and} \hskip 2mm
{\cal L}_{\tilde{S}/\tilde{G}}= pr(\tilde{K})
\]
if and only if $\tilde{K}=pr(\tilde{K})\tilde{\cap}{\cal
L}_{\tilde{G},m}$ and the fuzzy controllability condition of
$\tilde{K}$ with respect to $\tilde{G}$ and $\tilde{\Sigma}_{uc}$
holds, i.e., Eq. (9) holds.

{\it Proof:} The proof of sufficiency is similar to that of
Theorem 1. Set $\tilde{S}:
\tilde{\Sigma}^{*}\rightarrow {\cal F}(\tilde{\Sigma})$ as Eq. (10): For
any $\tilde{s}\in\tilde{\Sigma}^{*}$ and any
$\tilde{\sigma}\in\tilde{\Sigma}$,
\[
\tilde{S}(\tilde{s})(\tilde{\sigma})=\left\{\begin{array}{ll}
\min\left\{ \tilde{\Sigma}_{uc}(\tilde{\sigma}),{\cal
L}_{\tilde{G}}(\tilde{s}\tilde{\sigma})\right\}, & {\rm if}\hskip
2mm \tilde{\Sigma}_{uc}(\tilde{\sigma})\geq
pr(\tilde{K})(\tilde{s}\tilde{\sigma}),\\
pr(\tilde{K})(\tilde{s}\tilde{\sigma}),& {\rm otherwise}.
\end{array}
\right.
\]
Then with the same process as Theorem 1 we can prove that ${\cal
L}_{\tilde{S}/\tilde{G}}= pr(\tilde{K})$. Furthermore, since
$\tilde{K}= pr(\tilde{K})\tilde{\cap} {\cal L}_{\tilde{G},m}$, we
also have that
\[
{\cal L}_{\tilde{S}/\tilde{G},m}=  {\cal
L}_{\tilde{S}/\tilde{G}}\tilde{\cap} {\cal L}_{\tilde{G},m}=
pr(\tilde{K} )\tilde{\cap} {\cal L}_{\tilde{G},m}= \tilde{K},
\]
by which and ${\cal L}_{\tilde{S}/\tilde{G}}=
 pr(\tilde{K})$ we obtain $pr({\cal L}_{\tilde{S}/\tilde{G},m})={\cal
L}_{\tilde{S}/\tilde{G}}$, i.e., $\tilde{S}$ is nonblocking.

On the other hand, given that there exists an nonblocking
supervisory $\tilde{S}$ such that
 \[
{\cal L}_{\tilde{S}/\tilde{G},m} = \tilde{K}    \hskip 2mm {\rm
and} \hskip 2mm {\cal L}_{\tilde{S}/\tilde{G}} = pr(\tilde{K}),
\]
then with the definition of $ {\cal L}_{\tilde{S}/\tilde{G},m}$ we
have that
\[
\tilde{K}= {\cal L}_{\tilde{S}/\tilde{G},m} = {\cal L}_{\tilde{S}/\tilde{G}}\tilde{\cap}
{\cal L}_{\tilde{G},m} =pr(\tilde{K})\tilde{\cap} {\cal L}_{\tilde{G},m}.
\]
As well, the proof of fuzzy controllability condition is really
analogous to that of Theorem 1. Therefore, we  complete the proof
of this theorem. $\Box$

{\it Remark 6.} In Theorem 2, condition $pr(\tilde{K})\subseteq
{\cal L}_{\tilde{G},m}$ is necessary since $\tilde{K}\subseteq
{\cal L}_{\tilde{G},m}$ does not always result in
$pr(\tilde{K})\subseteq {\cal L}_{\tilde{G},m}$. (In Theorem 1
$pr(\tilde{K})\subseteq {\cal L}_{\tilde{G}}$ is not prerequisite,
because it was shown that $\tilde{K}\subseteq {\cal
L}_{\tilde{G}}$ deduces $pr(\tilde{K})\subseteq {\cal
L}_{\tilde{G}}$.)

{\it Example 7.} Let a fuzzy DES be modeled by a fuzzy finite
automaton $\tilde{G}=(\tilde{Q}, \tilde{\Sigma},
\tilde{\delta},\tilde{q}_{0},\tilde{Q}_{m})$ where
$\tilde{\Sigma}=\{\tilde{a}, \tilde{b}, \tilde{c}\}$.  The fuzzy
languages ${\cal L}_{\tilde{G}}$ and ${\cal L}_{\tilde{G},m}$ over
$\tilde{\Sigma}$ generated and marked by the fuzzy finite
automaton $\tilde{G}$ are the same as those in Example 6, that is,
${\cal L}_{\tilde{G}}(\epsilon)= 1$, $ {\cal
L}_{\tilde{G}}(\tilde{a})= {\cal L}_{\tilde{G}}
(\tilde{a}\tilde{b})= {\cal
L}_{\tilde{G}}(\tilde{a}\tilde{b}\tilde{c})= 0.8$, and ${\cal
L}_{\tilde{G}}(\tilde{s})=0$ for the other $s\in
\tilde{\Sigma}^{*}$; ${\cal L}_{\tilde{G},m}(\epsilon)= 1$, ${\cal
L}_{\tilde{G},m}( \tilde{a}\tilde{b} )= 0.8$, and ${\cal
L}_{\tilde{G},m}(\tilde{s})=0$ for the other $s\in
\tilde{\Sigma}^{*}$. Assume that a fuzzy set of control
specifications $ \tilde{K} $ is defined as: $\tilde{K}
(\epsilon)=1$, $\tilde{K} ( \tilde{a}\tilde{b} )=0.8$, and
$\tilde{K} (\tilde{s})=0$ for the other
$\tilde{s}\in\tilde{\Sigma}^{*}$. Clearly, $pr(\tilde{K})\subseteq
{\cal L}_{\tilde{G},m}$ and
$\tilde{K}=pr(\tilde{K})\tilde{\cap}{\cal L}_{\tilde{G},m}$. Also,
it is readily checked that the fuzzy controllability condition of
$\tilde{K}$ with respect to $\tilde{G}$ and $\tilde{\Sigma}_{uc}$
holds. Therefore, by means of Theorem 2 there is a supervisor
$\tilde{S}$ such that the supervised fuzzy control system ${\cal
L}_{\tilde{S}/\tilde{G}}$ satisfies
\[
{\cal L}_{\tilde{S}/\tilde{G},m}= \tilde{K}    \hskip 2mm {\rm and} \hskip 2mm
{\cal L}_{\tilde{S}/\tilde{G}}= pr(\tilde{K}).
\]
From the proof of Theorem 2 we know that supervisor $\tilde{S}:
\tilde{\Sigma}^{*}\rightarrow {\cal F}(\tilde{\Sigma})$ is defined
as Eq. (10). More concretely, in terms of Eq. (10) we have that $
\tilde{S}(\epsilon)(\tilde{a})=
\tilde{S}(\tilde{a})(\tilde{b})=0.8$, and 0 for the other cases.
It is seen that the deduced $\tilde{S}$ coincides exactly with
that in Example 6. \hskip 5mm $\Box$

Similarly to Corollary 1, with Theorem 2 we have:

{\it Corollary 2.}  Let a fuzzy DES be modeled by fuzzy finite
automaton (max-min or max-product)
$\tilde{G}=(\tilde{Q},\tilde{\Sigma},\tilde{\delta},\tilde{q}_{0},\tilde{Q}_{m})$.
 Let $n$ be any given positive integer.
Suppose fuzzy uncontrollable subset $\tilde{\Sigma}_{uc}\in {\cal
F}(\tilde{\Sigma})$, and fuzzy legal subset $\tilde{K}\in {\cal
F}( \tilde{\Sigma}^{*} )$ that satisfies: $\tilde{K}\subseteq_{n} {\cal
L}_{\tilde{G}}$, $pr(\tilde{K})\subseteq_{n} {\cal
L}_{\tilde{G},m}$, and $\tilde{K}(\epsilon)=1$. Then there exists a nonblocking supervisor
$\tilde{S}$ for $\tilde{G}$ such that $\tilde{S}$ satisfies the fuzzy
$n$-admissibility condition Eq. (6) and
\[
{\cal L}_{\tilde{S}/\tilde{G},m}=_{n} \tilde{K}    \hskip 2mm {\rm and} \hskip 2mm
{\cal L}_{\tilde{S}/\tilde{G}}=_{n} pr(\tilde{K})
\]
if and only if $\tilde{K}=_{n}pr(\tilde{K})\tilde{\cap}{\cal
L}_{\tilde{G},m}$ and the fuzzy $n$-controllability condition of
$\tilde{K}$ with respect to $\tilde{G}$ and $\tilde{\Sigma}_{uc}$
holds, i.e., Eq. (13) holds.

{\it Proof:} It is the same process as Theorem 2 by restricting the length of $\tilde{s}$ with $|\tilde{s}|\leq n$. \hskip 5mm $\Box$

\section*{IV. Properties of Controllability of Fuzzy DESs}

In this section, we deal with a number of basic properties
concerning supervisory controllability in fuzzy DESs.

{\it Definition 1.} Let $\tilde{K}$ and $\tilde{M}$ be fuzzy
languages over set $\tilde{\Sigma}$ of fuzzy events, and
$pr(\tilde{M})=\tilde{M}$. Suppose that $\tilde{\Sigma}_{uc}\in
{\cal F}(\tilde{\Sigma})$ denotes a fuzzy subset of
uncontrollable events. Then $\tilde{K}$ is said to be {\it
controllable with respect to $\tilde{M}$ and
$\tilde{\Sigma}_{uc}$} if for any $\tilde{s}\in
\tilde{\Sigma}^{*}$ and any $\tilde{\sigma}\in \tilde{\Sigma}$,
\begin{equation}
\min\left\{ pr(\tilde{K})(\tilde{s}), \tilde{\Sigma}_{uc}(\tilde{\sigma}),\tilde{M}( \tilde{s}\tilde{\sigma} )\right\}
\leq  pr(\tilde{K})(\tilde{s}\tilde{\sigma}).
\end{equation}

Intuitively, Eq. (19) means that the degree to which string
$\tilde{s}$ belongs to the prefix-closure of fuzzy language
$\tilde{K}$ and fuzzy event string $\tilde{\sigma}\tilde{s} $ also
belongs to fuzzy language $\tilde{M}$ together with $
\tilde{\sigma} $ being uncontrollable is not bigger than the
possibility of the string $\tilde{s}\tilde{\sigma}$ pertaining to
the prefix-closure of $\tilde{K}$. For convenience, we denote by
$C(\tilde{M},\tilde{\Sigma}_{uc})$ the set of all those being
controllable respect to $\tilde{M}$ and $\tilde{\Sigma}_{uc}$,
that is,
\[
C(\tilde{M},\tilde{\Sigma}_{uc})=\{\tilde{L}\in {\cal
F}(\tilde{\Sigma}^{*}):\tilde{L} \hskip 2mm {\rm is}\hskip 2mm
{\rm controllable}\hskip 2mm {\rm with} \hskip 2mm {\rm
respect}\hskip 2mm {\rm to}\hskip 2mm \tilde{M}\hskip 2mm {\rm
and}\hskip 2mm \tilde{\Sigma}_{uc}\}.
\]
Clearly, when $\tilde{M}$ is generated by some fuzzy finite
automaton $\tilde{G}$, i.e., $\tilde{M}={\cal L}_{\tilde{G}}$,
then Eq. (19) is the  same as the fuzzy controllable condition
described by Eq. (9) in Theorem 1. Also, it is easy to see that
$\tilde{K}\in C(\tilde{M},\tilde{\Sigma}_{uc})$ implies
$pr(\tilde{K})\in C(\tilde{M},\tilde{\Sigma}_{uc})$.

{\it Proposition 1.} Let $\tilde{M}\in {\cal
F}(\tilde{\Sigma}^{*})$ and  $ \tilde{\Sigma}_{uc}\in {\cal
F}(\tilde{\Sigma}) $ be the same as those in Definition 1. Suppose
that $\tilde{K}_{1},\tilde{K}_{2}$ are fuzzy languages over set
$\tilde{\Sigma}$ of fuzzy events.  Then:

(i) If  $\tilde{K}_{1},\tilde{K}_{2}\in
C(\tilde{M},\tilde{\Sigma}_{uc})$, then so is $
\tilde{K}_{1}\tilde{\cup}\tilde{K}_{2}$, where $\tilde{\cup}$ is
Zadeh fuzzy OR operator, that is,
$(\tilde{A}\tilde{\cup}\tilde{B})(x)=\max\{\tilde{A}(x),\tilde{B}(x)\}$.

(ii) If $pr(\tilde{K}_{1})\tilde{\cap}pr(\tilde{K}_{2})= pr(
\tilde{K}_{1}\tilde{\cap}\tilde{K}_{2} )$, and $\tilde{K}_{1}$ and
$\tilde{K}_{2}\in C(\tilde{M},\tilde{\Sigma}_{uc})$, then
$\tilde{K}_{1}\tilde{\cap}\tilde{K}_{2}\in
C(\tilde{M},\tilde{\Sigma}_{uc}) $, too.

(iii) If $pr(\tilde{K}_{i})=\tilde{K}_{i}$, $i=1,2$, and both
$\tilde{K}_{1}$ and $\tilde{K}_{2}\in
C(\tilde{M},\tilde{\Sigma}_{uc})$, then $pr(
\tilde{K}_{1}\tilde{\cap}\tilde{K}_{2})=\tilde{K}_{1}\tilde{\cap}\tilde{K}_{2}$
and $\tilde{K}_{1}\tilde{\cap}\tilde{K}_{2}\in
C(\tilde{M},\tilde{\Sigma}_{uc})$.

{\it Proof:} (i) For any $\tilde{s}\in \tilde{\Sigma}^{*}$ and any $\tilde{\sigma}\in
\tilde{\Sigma}$, we have
\begin{eqnarray*}
&&\min\left\{
pr(\tilde{K}_{1}\tilde{\cup}\tilde{K}_{2})(\tilde{s}),
\tilde{\Sigma}_{uc}(\tilde{\sigma}),
\tilde{M}(\tilde{s}\tilde{\sigma})\right\}\\
&=&\min\left\{\max\left\{
pr(\tilde{K}_{1})(\tilde{s}),pr(\tilde{K}_{2})(\tilde{s})\right\},\tilde{\Sigma}_{uc}(\tilde{\sigma}),
\tilde{M}(\tilde{s}\tilde{\sigma})\right\}\\
&=&\max\left\{
\min\left\{pr(\tilde{K}_{1})(\tilde{s}),\tilde{\Sigma}_{uc}(\tilde{\sigma}),
\tilde{M}(\tilde{s}\tilde{\sigma})\right\}, \min\left\{
pr(\tilde{K}_{1})(\tilde{s}),\tilde{\Sigma}_{uc}(\tilde{\sigma}),
\tilde{M}( \tilde{s}\tilde{\sigma} )\right\}\right\}\\
&\leq&\max\left\{ pr(\tilde{K}_{1})(\tilde{s}\tilde{\sigma}), pr(\tilde{K}_{2}) (\tilde{s}\tilde{\sigma})\right\}\\
&=&pr(\tilde{K}_{1}\tilde{\cup}\tilde{K}_{2})(\tilde{s}\tilde{\sigma}).
\end{eqnarray*}

(ii) For any $\tilde{s}\in \tilde{\Sigma}^{*}$ and any $\tilde{\sigma}\in
\tilde{\Sigma}$, we obtain that
\begin{eqnarray*}
&&\min\left\{
pr(\tilde{K}_{1}\tilde{\cap}\tilde{K}_{2})(\tilde{s}),
\tilde{\Sigma}_{uc}(\tilde{\sigma}),
\tilde{M}(\tilde{s}\tilde{\sigma})\right\}\\
&\leq&\min\left\{\min\left\{
pr(\tilde{K}_{1})(\tilde{s}),pr(\tilde{K}_{2})(\tilde{s})\right\},\tilde{\Sigma}_{uc}(\tilde{\sigma}),
\tilde{M}(\tilde{s}\tilde{\sigma})\right\}\\
&=&\min\left\{
\min\left\{pr(\tilde{K}_{1})(\tilde{s}),\tilde{\Sigma}_{uc}(\tilde{\sigma}),
\tilde{M}(\tilde{s}\tilde{\sigma})\right\}, \min\left\{
pr(\tilde{K}_{1})(\tilde{s}),\tilde{\Sigma}_{uc}(\tilde{\sigma}),
\tilde{M}( \tilde{s}\tilde{\sigma} )\right\}\right\}\\
&\leq&\min\left\{ pr(\tilde{K}_{1})(\tilde{s}\tilde{\sigma}), pr(\tilde{K}_{2}) (\tilde{s}\tilde{\sigma})\right\}\\
&=& pr(\tilde{K}_{1}\tilde{\cap}\tilde{K}_{2})(\tilde{s}\tilde{\sigma}).
\end{eqnarray*}

(iii) For any $\tilde{s}\in \tilde{\Sigma}^{*}$ and any $\tilde{\sigma}\in
\tilde{\Sigma}$, we have that
\begin{eqnarray*}
 pr ( \tilde{K}_{1}\tilde{\cap}\tilde{K}_{2})(\tilde{s})&=&
\sup_{\tilde{s}\in pr(\tilde{t})} (\tilde{K}_{1}\tilde{\cap}\tilde{K}_{2})(\tilde{t})\\
&=& \sup_{\tilde{s}\in pr(\tilde{t})}\min\{ \tilde{K}_{1}(\tilde{t}),\tilde{K}_{2}(\tilde{t})\}\\
&\leq&\min \left\{ \sup_{\tilde{s}\in pr(\tilde{t})}\tilde{K}_{1}(\tilde{t}),  \sup_{\tilde{s}\in pr(\tilde{t})}\tilde{K}_{2}(\tilde{t})\right\}\\
&=&\min\left\{ pr(\tilde{K}_{1})(\tilde{s}),pr(\tilde{K}_{2})(\tilde{s})\right\}\\
&=&\min\left\{ \tilde{K}_{1}(\tilde{s}),\tilde{K}_{2}(\tilde{s})\right\}\\
&=&(\tilde{K}_{1}\tilde{\cap}\tilde{K}_{2})(\tilde{s})\\
&\leq& pr ( \tilde{K}_{1}\tilde{\cap}\tilde{K}_{2})(\tilde{s}).
\end{eqnarray*}
The rest of the proof is similar to (ii). $\Box$

Given a fuzzy language $\tilde{K}$ over set of fuzzy events
$\tilde{\Sigma}$, we hope to find the largest fuzzy sublanguage
and the smallest prefix-closed fuzzy controllable superlanguage of
$\tilde{K}$, such that these languages are
controllable with respect to $\tilde{M}$ and
$\tilde{\Sigma}_{uc}$. To this end, we denote

$
{\cal K}(\tilde{K})^{<} =\{\tilde{L}\subseteq \tilde{K}:
\tilde{L}\in C(\tilde{M},\tilde{\Sigma}_{uc})\},
$

$
{\cal K}(\tilde{K})^{>} =\{\tilde{L}\in {\cal
F}(\tilde{\Sigma}^{*}):\tilde{K}\subseteq\tilde{L}\subseteq\tilde{M}\hskip
2mm {\rm and} \hskip 2mm pr(\tilde{L})=\tilde{L}\hskip 2mm {\rm
and}\hskip 2mm  \tilde{L}\in C(\tilde{M},\tilde{\Sigma}_{uc})\},
$

$
\tilde{K}^{<} =\tilde{\bigcup}_{\tilde{K}\in {\cal
K}(\tilde{K})^{<}}\tilde{L}, \hskip 2mm {\rm and} \hskip 2mm
\tilde{K}^{>} =\tilde{\bigcap}_{\tilde{K}\in {\cal
K}(\tilde{K})^{>}}\tilde{L}.
$

{\it Lemma 1.} Let $\tilde{M}\in {\cal F}(\tilde{\Sigma}^{*})$ and
$\tilde{\Sigma}_{uc}\in {\cal F}(\tilde{\Sigma}) $ be the same as
those in Definition 1. Suppose $\tilde{K}\in {\cal
F}(\tilde{\Sigma}^{*})$. Then:

 (i)
$\tilde{K}^{<}\in {\cal K}(\tilde{K})^{<}$.

(ii) $\tilde{K}^{>}\in {\cal K}(\tilde{K})^{>}$.

{\it Proof:} (i) With the proof of Proposition 1 (i) it suffices
to show that $ pr(\tilde{\cup}_{i\in I}\tilde{L}_{i}) =
\tilde{\cup}_{i\in I} pr(\tilde{L}_{i})$, where $I$ denotes an
index set (perhaps infinite). For any
$\tilde{s}\in\tilde{\Sigma}^{*}$, we have
\begin{eqnarray*}
pr( \tilde{\cup}_{i\in I}\tilde{L}_{i} )(\tilde{s})&=& \sup_{\tilde{s}\in pr(\tilde{t})} ( \tilde{\cup}_{i\in I}\tilde{L}_{i})(\tilde{t})\\
&=&\sup_{\tilde{s}\in pr(\tilde{t})} \sup_{i\in I}\tilde{L}_{i}(\tilde{t})\\
&=& \sup_{i\in I} \sup_{\tilde{s}\in pr(\tilde{t})}\tilde{L}_{i}(\tilde{t})\\
&=& \sup_{i\in I} pr(\tilde{L}_{i})(\tilde{s})\\
&=&(\tilde{\cup}_{i\in I} pr(\tilde{L}_{i}))(\tilde{s}).
\end{eqnarray*}
 (ii)  Since $pr(\tilde{L}_{i})= \tilde{L}_{i} $ for each $\tilde{L}_{i}\in \tilde{{\cal K}}(\tilde{K})^{>} $, we have that for any
$\tilde{s}\in\tilde{\Sigma}^{*}$,
\begin{eqnarray*}
pr( \tilde{K}^{>} )(\tilde{s})&=&pr( \tilde{\cap}_{\tilde{L}_{i} \in {\cal K}(\tilde{K})^{>}}\tilde{L}_{i} )(\tilde{s})\\
&=& \sup_{\tilde{s}\in pr(\tilde{t})}(\tilde{\cap}_{ \tilde{L}_{i} \in {\cal K}(\tilde{K})^{>}} \tilde{L}_{i} )(\tilde{t})\\
&=& \sup_{\tilde{s}\in pr(\tilde{t})}\inf_{\tilde{L}_{i} \in {\cal K}(\tilde{K})^{>} }\tilde{L}_{i} (\tilde{t})\\
&\leq& \inf_{\tilde{L}_{i} \in {\cal K}(\tilde{K})^{>} } \sup_{\tilde{s}\in pr(\tilde{t})} \tilde{L}_{i} (\tilde{t})\\
&=& \inf_{\tilde{L}_{i} \in {\cal K}(\tilde{K})^{>} } pr( \tilde{L}_{i} ) (\tilde{s})\\
&=&\inf_{\tilde{L}_{i} \in {\cal K}(\tilde{K})^{>} }  \tilde{L}_{i}  (\tilde{s})\\
&=&(\tilde{\cap}_{\tilde{L}_{i} \in {\cal K}(\tilde{K})^{>} } \tilde{L}_{i}  )(\tilde{s})\\
&=& \tilde{K}^{>} (\tilde{s})\\
&\leq& pr(\tilde{K}^{>}) (\tilde{s}),
\end{eqnarray*}
and therefore $pr(\tilde{K}^{>})=\tilde{K}^{>}$. The proof of
$\tilde{K}^{>}\in C(\tilde{M},\tilde{\Sigma}_{uc})$ is similar to
Proposition 1 (ii). $\Box$

Clearly, when $\tilde{K}\in C(\tilde{M},\tilde{\Sigma}_{uc})$, we
have $\tilde{K}^{<}=\tilde{K}$. Concerning  $\tilde{K}^{<}$ and
 $\tilde{K}^{>}$, we further have the following properties.

{\it Proposition 2.} Let $\tilde{M}\in {\cal
F}(\tilde{\Sigma}^{*})$ and  $ \tilde{\Sigma}_{uc}\in {\cal
F}(\tilde{\Sigma}) $ be the same as those in Definition 1. Suppose
that $\tilde{K}_{1},\tilde{K}_{2}$ are fuzzy languages over set
$\tilde{\Sigma}$ of fuzzy events.  Then:

(i) If $\tilde{K}$ satisfies $pr(\tilde{K})=\tilde{K}$, then so is
$ \tilde{K}^{<} $, i.e., $pr(\tilde{K}^{<})=\tilde{K}^{<}$.

(ii) If $ \tilde{K}_{1} \subseteq \tilde{K}_{2}$, then $ \tilde{K}_{1}^{<} \subseteq \tilde{K}_{2}^{<}$.

(iii) $ ( \tilde{K}_{1}\tilde{\cap} \tilde{K}_{2} )^{<}\subseteq \tilde{K}_{1}^{<}\tilde{\cap}\tilde{K}_{2}^{<}$.

(iv) $ ( \tilde{K}_{1}\tilde{\cap} \tilde{K}_{2} )^{<}=
(\tilde{K}_{1}^{<}\tilde{\cap}\tilde{K}_{2}^{<})^{<}$.

(v) If $pr(  \tilde{K}_{1}^{<}\tilde{\cap}\tilde{K}_{2}^{<} )=
pr(\tilde{K}_{1}^{<})\tilde{\cap} pr(\tilde{K}_{2}^{<})$, then
$
( \tilde{K}_{1}\tilde{\cap} \tilde{K}_{2} )^{<}=\tilde{K}_{1}^{<}\tilde{\cap}\tilde{K}_{2}^{<}.
$

(vi) $\tilde{K}_{1}^{<}\tilde{\cup}\tilde{K}_{2}^{<}\subseteq (
\tilde{K}_{1}\tilde{\cup} \tilde{K}_{2} )^{<}$.

{\it Proof:} (i) With Lemma 1 (i) we know that $\tilde{K}^{<}\in
\tilde{{\cal K}}(\tilde{K})^{<}$, and thus
\[
pr(\tilde{K}^{<})\subseteq pr(\tilde{K})=\tilde{K}\subseteq
pr(\tilde{K}),
\]
which results in $pr(\tilde{K}^{<})=\tilde{K}^{<}$.

(ii) Straightforward.

(iii) It follows clearly from the definition of $\tilde{K}^{<}$.

(iv) Since
$\tilde{K}_{1}^{<}\tilde{\cap}\tilde{K}_{2}^{<}\subseteq\tilde{K}_{1}\tilde{\cap}\tilde{K}_{2}$
from Lemma 1 (i), with (ii) we obtain that
$(\tilde{K}_{1}^{<}\tilde{\cap}\tilde{K}_{2}^{<})^{<}\subseteq(\tilde{K}_{1}\tilde{\cap}\tilde{K}_{2})^{<}$.
On the other hand, for any $\tilde{L}\in {\cal
K}(\tilde{K}_{1}\tilde{\cap}\tilde{K}_{2})^{<}$, then
$\tilde{L}\in C(\tilde{M},\tilde{\Sigma}_{uc})$ and
$\tilde{L}\subseteq\tilde{K}_{1}\tilde{\cap}\tilde{K}_{2}$. Thus
$\tilde{L}\subseteq \tilde{K}_{1}$ and $\tilde{L}\subseteq
\tilde{K}_{2}$. Since $\tilde{K}_{1}^{<}\subseteq\tilde{K}_{1}$
and $\tilde{K}_{2}^{<}\subseteq\tilde{K}_{2}$ by means of Lemma 1
(i), with the definition of $\tilde{K}^{<}$ we obtain that
$\tilde{L}\subseteq\tilde{K}_{1}^{<}$ and
$\tilde{L}\subseteq\tilde{K}_{2}^{<}$, and thus
$\tilde{L}\subseteq\tilde{K}_{1}^{<}\tilde{\cap}\tilde{K}_{2}^{<}$.
Therefore, for any $\tilde{s}\in\tilde{\Sigma}^{*}$,
\begin{eqnarray*}
(\tilde{K}_{1}\tilde{\cap}\tilde{K}_{2})^{<}(\tilde{s})&=&\sup\{\tilde{L}(\tilde{s}):\tilde{L}\in
{\cal K}(\tilde{K}_{1}\tilde{\cap}\tilde{K}_{2})^{<}\}\\
&\leq&
\sup\{\tilde{L}(\tilde{s}):\tilde{L}\subseteq\tilde{K}_{1}^{<}\tilde{\cap}\tilde{K}_{2}^{<},\tilde{L}\in
C(\tilde{M},\tilde{\Sigma}_{uc})\}\\
&=&(\tilde{K}_{1}^{<}\tilde{\cap}\tilde{K}_{2}^{<})^{<}(\tilde{s}).
\end{eqnarray*}
As a result,
$(\tilde{K}_{1}\tilde{\cap}\tilde{K}_{2})^{<}\subseteq
(\tilde{K}_{1}^{<}\tilde{\cap}\tilde{K}_{2}^{<})^{<}$, and (iv) is
thus proved.

(v) From Proposition 1 (ii) and Lemma 1 (i) it follows that
$\tilde{K}_{1}^{<}\tilde{\cap}\tilde{K}_{2}^{<}\in
C(\tilde{M},\tilde{\Sigma}_{uc})$. Thus, with (iv) just verified
we have that
\[
\tilde{K}_{1}^{<}\tilde{\cap}\tilde{K}_{2}^{<}=(\tilde{K}_{1}^{<}\tilde{\cap}\tilde{K}_{2}^{<})^{<}=(\tilde{K}_{1}\tilde{\cap}\tilde{K}_{2})^{<}.
\]

(vi) For any $\tilde{s}\in\tilde{\Sigma}^{*}$, in light of the
definition of ${\cal K}(\tilde{K})^{<}$, we have that
\begin{eqnarray*}
(\tilde{K}_{1}^{<}\tilde{\cup}\tilde{K}_{2}^{<})(\tilde{s})&=&\max\{\tilde{K}_{1}^{<}(\tilde{s}),\tilde{K}_{2}^{<}(\tilde{s})\}\\
&=&\max\left\{\sup_{\tilde{L}_{1}\in {\cal
K}(\tilde{K}_{1})^{<}}\tilde{L}_{1}(\tilde{s}),\sup_{\tilde{L}_{2}\in
{\cal K}(\tilde{K}_{2})^{<}}\tilde{L}_{2}(\tilde{s})\right\}\\
&\leq&\sup_{\tilde{L}\in {\cal
K}(\tilde{K}_{1}\tilde{\cup}\tilde{K}_{2})^{<}}\tilde{L}(\tilde{s}).\hskip
4mm \Box
\end{eqnarray*}

{\it Proposition 3.} Let $\tilde{M}\in {\cal
F}(\tilde{\Sigma}^{*})$ and  $ \tilde{\Sigma}_{uc}\in {\cal
F}(\tilde{\Sigma}) $ be the same as those in Definition 1. Suppose
that $\tilde{K}_{1},\tilde{K}_{2}$ are fuzzy languages over set
$\tilde{\Sigma}$ of fuzzy events. Then:

(i) If $\tilde{K}\in C(\tilde{M},\tilde{\Sigma}_{uc})$, then
$\tilde{K}^{>}=pr(\tilde{K})$.

(ii) If $\tilde{K}_{1}\subseteq \tilde{K}_{2}$, then
$\tilde{K}_{1}^{>}\subseteq \tilde{K}_{2}^{>}$.

(iii) $(\tilde{K}_{1}\tilde{\cap}\tilde{K}_{2})^{>}\subseteq
(\tilde{K}_{1}^{>}\tilde{\cap}\tilde{K}_{2}^{>})^{>}$.

(iv) $ ( \tilde{K}_{1}\tilde{\cup} \tilde{K}_{2} )^{>}=
\tilde{K}_{1}^{>}\tilde{\cup}\tilde{K}_{2}^{>}$.

{\it Proof:} (i) If $\tilde{K}\in
C(\tilde{M},\tilde{\Sigma}_{uc})$, then also $pr(\tilde{K})\in
C(\tilde{M},\tilde{\Sigma}_{uc})$. Together with
$\tilde{K}\subseteq pr(\tilde{K})$ and
$pr(pr(\tilde{K}))=pr(\tilde{K})$ we obtain that $pr(\tilde{K})\in
{\cal K}(\tilde{K})^{>}$ since $\tilde{M}=pr(\tilde{M})$. Thus,
$\tilde{K}^{>}\subseteq pr(\tilde{K})$. Contrarily, for any
$\tilde{L}\in {\cal K}(\tilde{K})^{>}$, then
$pr(\tilde{L})=\tilde{L}$ and  we thus have
\[
pr(\tilde{K})\subseteq pr(\tilde{L})=\tilde{L}.
\]
Therefore $pr(\tilde{K})\subseteq \tilde{\bigcap}_{\tilde{L}\in
{\cal K}(\tilde{K})^{>}}\tilde{L}=\tilde{K}^{>}$.

The demonstrations of (ii) and (iii) are clear, so, we here omit
the details.

(iv) Due to $\tilde{K}_{1}\subseteq \tilde{K}_{1}\tilde{\cup}
\tilde{K}_{2}$ and $\tilde{K}_{2}\subseteq
\tilde{K}_{1}\tilde{\cup} \tilde{K}_{2}$, we have
\[
{\cal K}(\tilde{K}_{1}\tilde{\cup} \tilde{K}_{2})^{>}\subseteq
{\cal K}(\tilde{K}_{1})^{>}\tilde{\cup} {\cal
K}(\tilde{K}_{2})^{>}.
\]
Thus,
\[
\tilde{K}_{1}^{>}\tilde{\cup}\tilde{K}_{2}^{>}\subseteq
(\tilde{K}_{1}\tilde{\cup} \tilde{K}_{2})^{>}.
\]
On the other hand, for any $\tilde{s}\in\tilde{\Sigma}^{*}$,
suppose that
\begin{equation}
(\tilde{K}_{1}\tilde{\cup}
\tilde{K}_{2})^{>}(\tilde{s})=\inf_{\tilde{L}\in {\cal
K}(\tilde{K}_{1}\tilde{\cup}
\tilde{K}_{2})^{>}}\tilde{L}(\tilde{s})>t
\end{equation}
for $t\in [0,1)$. Then we claim that
\begin{equation}
(\tilde{K}_{1}^{>}\tilde{\cup}\tilde{K}_{2}^{>})(\tilde{s})>t;
\end{equation}
otherwise, $\tilde{K}_{1}^{>}(\tilde{s})\leq t$ and
$\tilde{K}_{2}^{>}(\tilde{s})\leq t$, from which together with the
definition of
$(\tilde{K}_{1}^{>}\tilde{\cup}\tilde{K}_{2}^{>})(\tilde{s})$,
i.e.,
\[
(\tilde{K}_{1}^{>}\tilde{\cup}\tilde{K}_{2}^{>})(\tilde{s})=\max\left\{\inf_{\tilde{L}_{1}\in
{\cal
K}(\tilde{K}_{1})^{>}}\tilde{L}_{1}(\tilde{s}),\inf_{\tilde{L}_{2}\in
{\cal K}(\tilde{K}_{2})^{>}}\tilde{L}_{2}(\tilde{s})\right\},
\]
it follows that for any $\varepsilon>0$, there exist
$\tilde{L}_{1}\in {\cal K}(\tilde{K}_{1})^{>}$ and
$\tilde{L}_{2}\in {\cal K}(\tilde{K}_{2})^{>}$ such that
$\tilde{L}_{1}(\tilde{s})<t+\varepsilon$ and
$\tilde{L}_{2}(\tilde{s})<t+\varepsilon$. Due to
$pr(\tilde{L}_{1}\tilde{\cup}\tilde{L}_{2})=pr(\tilde{L}_{1})\tilde{\cup}pr(\tilde{L}_{2})$
and Proposition 1 (i), it follows that
$\tilde{L}_{1}\tilde{\cup}\tilde{L}_{2}\in {\cal
K}(\tilde{L}_{1}\tilde{\cup}\tilde{L}_{2})^{>}$. Consequently,
\begin{eqnarray*}
(\tilde{L}_{1}\tilde{\cup}\tilde{L}_{2})^{>}(\tilde{s})&\leq&
(\tilde{L}_{1}\tilde{\cup}\tilde{L}_{2})(\tilde{s})\\
&=&\max\{\tilde{L}_{1}(\tilde{s}),\tilde{L}_{2}(\tilde{s})\}\\
&<&t+\varepsilon.
\end{eqnarray*}
Since $\varepsilon$ is arbitrary, we have
$(\tilde{L}_{1}\tilde{\cup}\tilde{L}_{2})^{>}(\tilde{s})\leq t$,
which contradicts the assumption of Eq. (20). Therefore, Eq. (21),
i.e., $(\tilde{L}_{1}\tilde{\cup}\tilde{L}_{2})^{>}(\tilde{s})> t$
holds true. As a consequence, $ (\tilde{K}_{1}\tilde{\cup}
\tilde{K}_{2})^{>}(\tilde{s})\leq
(\tilde{K}_{1}^{>}\tilde{\cup}\tilde{K}_{2}^{>})(\tilde{s}) $, and
therefore $ (\tilde{K}_{1}\tilde{\cup} \tilde{K}_{2})^{>}\subseteq
\tilde{K}_{1}^{>}\tilde{\cup}\tilde{K}_{2}^{>} $ is proved. $\Box$

\section*{V. Concluding Remarks}

Fuzzy DESs initiated by Lin and Ying [19,18] may conform more to
human's perception when coping with some real-world problems,
especially in biomedical applications and likely in traffic
systems (the changes of traffic light are controlled in terms of
the length of queues of vehicles, in which length is a fuzzy
concept). With a desire to provide certain foundation for the
applications of fuzzy DESs, in this paper we further developed
fuzzy DESs by dealing with supervisory control issue of fuzzy
DESs. The technical contributions are mainly as follows: (i) we
reformulated the parallel composition of crisp DESs, and defined
the parallel composition of fuzzy DESs that is equivalent to that
in [19]; max-min and max-product automata for modeling DESs are
considered; (ii) we dealt with a number of basic problems in
supervisory control of fuzzy DESs, and demonstrated
controllability theorem and nonblocking controllability theorem of
fuzzy DESs, and thus obtained the conditions for the existence of
supervisors in fuzzy DESs; (iii) after analyzing the complexity
for presenting a uniform criterion for testing the fuzzy
controllability condition of fuzzy DESs modeled by {\it
max-product} automata, we particularly presented in detail a
computing flow for checking the fuzzy controllability condition of
fuzzy DESs modeled by {\it max-min} automata, and by means of this
method  we can search for all possible fuzzy states reachable from
initial fuzzy state in max-min automata; as well, the fuzzy
$n$-controllability condition was introduced for some realistic
control problems; (iv) a number of examples serving to illustrate
the applications of the obtained methods and theorems were
described; (v) some basic properties related to supervisory
control of fuzzy DESs were discussed, which include the existence
of supremal fuzzy controllable sublanguage and infimal
prefix-closed fuzzy controllable superlanguage of a given fuzzy
noncontrollable sublanguage.

As was known [1], in classical DESs, under certain condition we
can check within finite number of steps whether the
controllability condition holds by utilizing the finiteness of
states in crisp finite automata, but the infiniteness of fuzzy
states in fuzzy DESs modeled by max-product automata gives rise to
considerable complexity for formulating a uniform fashion to check
the fuzzy controllability conditions. However, when  fuzzy DESs
are modeled by max-min automata, we have derived a uniform
criterion (computing flow) to check this condition. This is a very
important result in supervisory control of fuzzy DESs. Notably,
our methods also apply to testing supervisory control condition of
crisp DESs. Therefore, it has been seen that there are some
essential differences between fuzzy DESs and crisp DESs, and the
computing processes in fuzzy DESs are also more complicated. A
significant issue is to give some efficient algorithms in fuzzy
DESs such as constructing efficient algorithms to check the fuzzy
controllability conditions presented in Theorems 1 and 2, on the
basis of the computing flow described in this paper.

With the results obtained in Ref. [19] and this paper, a further
issue also worthy of consideration is to deal with supervisory
control of fuzzy DESs modeled via max-min automata under partial
observation, which includes how to establish controllability and
observability theorem of fuzzy DESs, and how to deal with
decentralized supervisory control of fuzzy DESs. Another important
modeling approach to fuzzy DESs, that is, fuzzy DESs modeled by
fuzzy Petri nets [15], has not been investigated, and we deem it a
significant research direction. Finally, it is worth indicating
that computing with words, as a methodology, advocated by Zadeh
and others [38,39,40,37,36], may play a useful role in the further
exploration of fuzzy DESs. These problems will be considered in
subsequent work.

\section*{Acknowledgments}

The author wishes to thank  L. O. Hall, and the Associate Editor,
as well as the anonymous reviewers for their invaluable
suggestions and comments that greatly helped to improve the
quality of this paper.

\end{document}